\documentclass[usenatbib,usegraphicx]{mn2e}

\renewcommand{\d}{\ensuremath{\mathrm{d}}}
\newcommand\ion[2]{#1$\;${\small\rmfamily{#2}}}%

\newcommand{\SN}{\ensuremath{\mathrm{S/N}}}
\newcommand{\sauron}{\texttt{SAURON}}

\newcommand{\plotone}[1]{%
  \includegraphics[width=\columnwidth]{#1}}
\newcommand{\plottwo}[2]{%
  \includegraphics[width=\columnwidth]{#1}
  \includegraphics[width=\columnwidth]{#2}}

\newcommand{\refsec}[1]{section~\ref{#1}}
\newcommand{\reffig}[1]{Fig.~\ref{#1}}

\title[Adaptive 2D-binning]{%
  Adaptive spatial binning of integral-field spectroscopic data using
  Voronoi tessellations}

\author[M.~Cappellari \& Y.~Copin]{%
  Michele Cappellari$^{1}$\thanks{European Space Agency external fellow.}
and
  Yannick Copin$^{1,2}$ \\
  $^{1}$ Leiden Observatory, Postbus 9513, 2300 RA Leiden, The Netherlands \\
  $^{2}$ Institut de physique nucl\'eaire de Lyon, 69222 Villeurbanne, France
}

\pagerange{\pageref{firstpage}--\pageref{lastpage}} \pubyear{2003}

\begin{document}
\label{firstpage}

% Abstract ------------------------------

\maketitle

\begin{abstract}
  We present new techniques to perform adaptive spatial binning of
  Integral-Field Spectroscopic (IFS) data to reach a chosen constant
  signal-to-noise ratio per bin. These methods are required for the
  proper analysis of IFS observations, but can also be used for
  standard photometric imagery or any other two-dimensional data.
  Various schemes are tested and compared by binning and extracting
  the stellar kinematics of the Sa galaxy NGC~2273 from spectra
  obtained with the panoramic IFS \sauron.
\end{abstract}

% Text ==============================

% Introduction ------------------------------

\section{Introduction}
\label{sec:intro}

Spatially resolved astronomical observations commonly span orders of
magnitude variations in the signal-to-noise ratio (\SN) across the
detector elements, and some spatial elements are often flawed by
insufficient \SN. For this reason, data are often locally averaged
together before analyzing them: the resulting \SN{} is significantly
better at the price of a degraded spatial resolution.

Two kinds of averaging processes are possible: \emph{smoothing} and
\emph{binning}. The smoothing scheme increases the local \SN{} by
somehow correlating neighboring data --- e.g., during a convolution
process --- while retaining the initial sampling, and therefore the
initial number of data points. On the other hand, the binning
technique locally groups and averages together data, resulting in a
sparser sampling and a decreased number of final data points. In both
cases, the number of \emph{independent} measurements is equivalently
decreased. While the smoothing technique might present the advantage
of simplicity --- see e.g., the very wide use of the `median
filtering' --- it is difficult to handle the statistics of the
correlations it introduces. As a consequence, smoothing is never used
for the quantitative analysis of one-dimensional (1D) spectra. When a
precise error management is needed, or when the actual number of data
points to be treated matters (e.g., for a heavy subsequent numerical
computation), binning is more suited.

Given the large variations in the \SN{} across the detector elements
--- e.g., from the inner to the outer parts of a galaxy CCD image ---
it is of interest to use an \emph{adaptive binning} scheme where the
size of the bin is adapted to the local \SN{}: bigger bins will be
applied in the low-\SN{} regions, while a higher resolution --- i.e.,
smaller bins --- will be retained in the high-\SN{} parts. A well
known example is galaxy photometry, where logarithmically spaced
radial bins are often adopted \citep[e.g.][]{jed87}: many more pixels
are used to compute the value of a galaxy profile at large radii than
in the center. However, these algorithms generally use an \emph{a
  priori} knowledge of the \SN{} distribution (e.g. the surface
brightness profile of the galaxy) to build the binning scheme.

Binning is essential in the case of spectroscopic observations of the
stellar kinematics. In fact the extraction of the line-of-sight
velocity distribution (LOSVD) generally involves nonlinear processes
and as a result a minimum \SN{} is \emph{required} for a reliable and
unbiased extraction of kinematical information from the spectra
\citep[e.g.][and references therein]{UGD}. Accordingly, binning is
invariably used to analyze 1D (e.g.\ long slit)
spectroscopic observations.  Developments with Integral-Field
Spectroscopy (IFS; e.g., \sauron{} on WHT, \texttt{VIMOS} on VLT,
\texttt{GMOS} on Gemini, etc.) require methods to perform spatial
binning of spectra in two dimensions (2D) too.

Little work has been done on the subject of fully adaptive 2D-binning.
\citet{sanders01} developed an algorithm for X-ray imaging data, but
the bins that their method produces can contain other bins and are not
compact. In the case of spectroscopic data it makes little sense to
bin together spectra coming from pixels that are not close enough to
each other and whose properties may differ considerably. Other schemes
have to be developed and are described in this paper. In
\refsec{sec:generic} we discuss the specific requirements of the
2D-binning problem. In \refsec{sec:quadtree} we present a solution
based on the Quadtree method, while in \refsec{sec:vt} we develop new
methods based on Voronoi tessellations. In \refsec{sec:optimal} we
present an optimal Voronoi 2D-binning algorithm and finally in
\refsec{sec:concl} we draw some conclusions.

\section{Formulation of the problem}
\label{sec:generic}

We tackle here the problem of binning in the spatial direction(s). In
what follows the term `pixel' refers to a given spatial element of the
dataset (sometimes called `spaxel' in the IFS community): it can be an
actual pixel of a CCD image, or a spectrum position along the slit of
a long-slit spectrograph or in the field of view of an IFS (e.g.\ a
lenslet or a fiber).

Each pixel $i$ has an associated signal $\mathcal{S}_{i}$ and its
corresponding noise $\mathcal{N}_{i}$. The pixel signal-to-noise ratio
is therefore $(\SN)_{i} = \mathcal{S}_{i}/\mathcal{N}_{i}$. Our
considerations do not depend on the details used to estimate these
quantities, which we assume to be known beforehand. In the case of
spectroscopy for instance, the quantity $\mathcal{S}_{i}$ (resp.
${\cal N}_{i}$) associated with a given spectrum $S_{i}(\lambda)$ can
be the signal (resp. the noise $N_i(\lambda)$) averaged over a given
spectral range $\Delta\lambda$:
\begin{equation}
  \mathcal{S}_{i} = \frac{1}{\Delta\lambda}\int_{\Delta\lambda}
    S_{i}(\lambda)\,\d\lambda,
\quad
  \mathcal{N}^2_{i} = \frac{1}{\Delta\lambda}\int_{\Delta\lambda}
    N^2_{i}(\lambda)\,\d\lambda.
\end{equation}
When two pixels are coadded the \SN{} of the resulting bin is computed
according to the standard formula:
\begin{equation}
(\SN)_{1+2} = \frac{\mathcal{S}_{1}+\mathcal{S}_{2}}
                   {\sqrt{\mathcal{N}^2_{1}+\mathcal{N}^2_{2}}}.
\label{eq:sn_sum}
\end{equation}
It is important to remember that the term `binning' will only refer
here to the averaging of observations taken at different positions on
the sky (i.e., different pixels), and \emph{not} along the spectral
direction.

The binning method in 1D is easy to implement: one only has to make
sure that all the bins are adjacent and that their \SN{} reach a
minimum value (or better that the scatter around the target \SN{} is
minimum).  This leads in practice to a unique binning solution (see a
practical algorithm in \refsec{sec:accretion}). In 2D (and higher
dimensions) the situation is more complex as the shape of the bin has
then to be taken into consideration. A good binning scheme has to
satisfy the following requirements:
\begin{description}
\item [\textbf{Topological requirement:}] the bins should properly
  tessellate the relevant region $\Omega$ of the sky plane, i.e.,
  create a partition of $\Omega$ without overlapping or holes. While
  this requirement is trivial to enforce in 1D, it is tricky to
  implement in higher dimensions, where the bin shapes vary;

\item [\textbf{Morphological requirement:}] the bin shape has to be as
  `compact' (or `round') as possible, so that the pixels in one bin
  are as close as possible to each other and can be associated with a
  well-defined spatial position. In this way the best spatial
  resolution is obtained along all directions;

\item [\textbf{Uniformity requirement:}] the scatter in the \SN{} of
  the bins, around a target value, should be as small as possible.
  While a minimum \SN{} is generally required, one does not want to
  sacrifice spatial resolution to increase the bin \SN{} even further.
\end{description}

\begin{figure}
  \plottwo{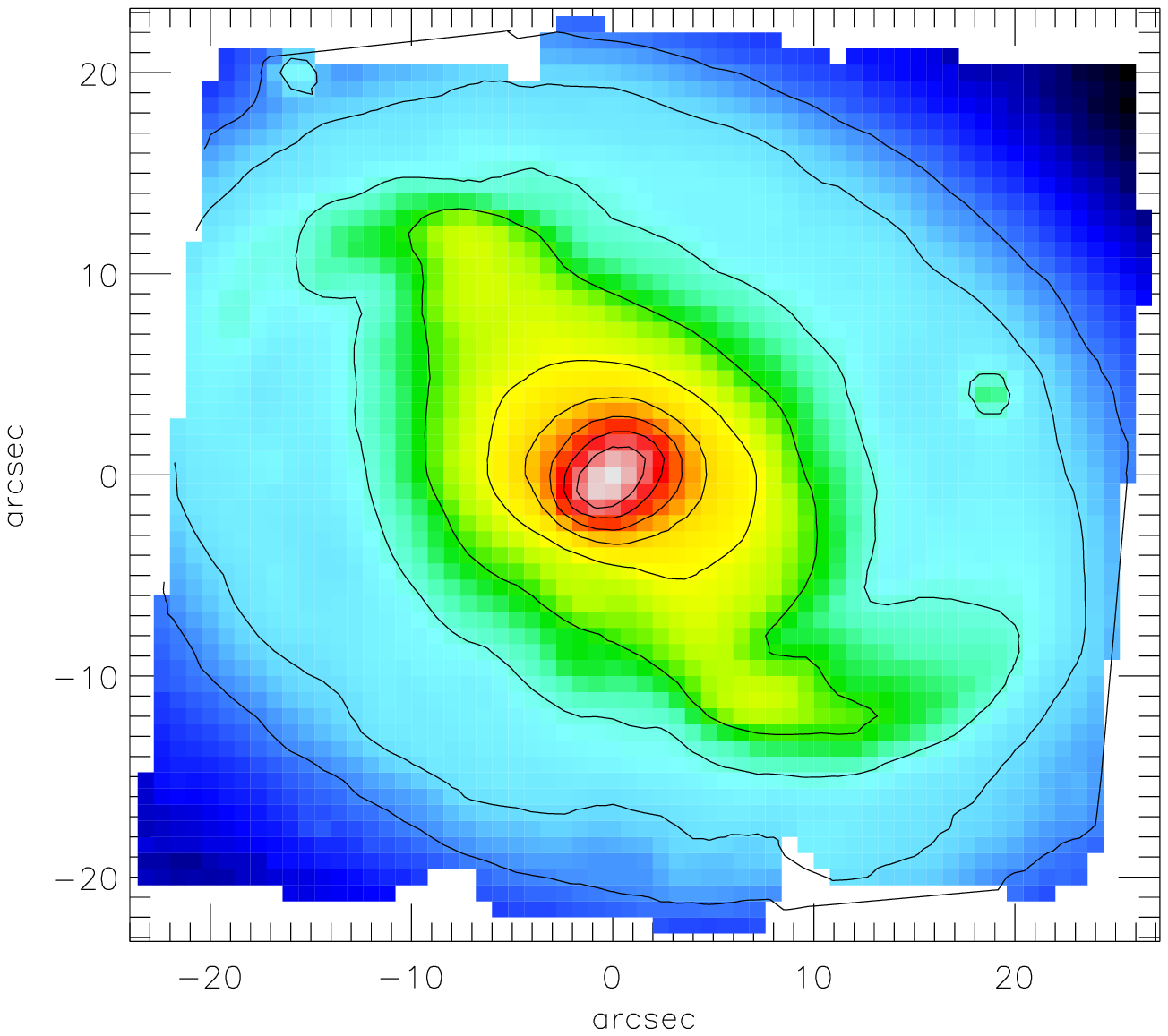}{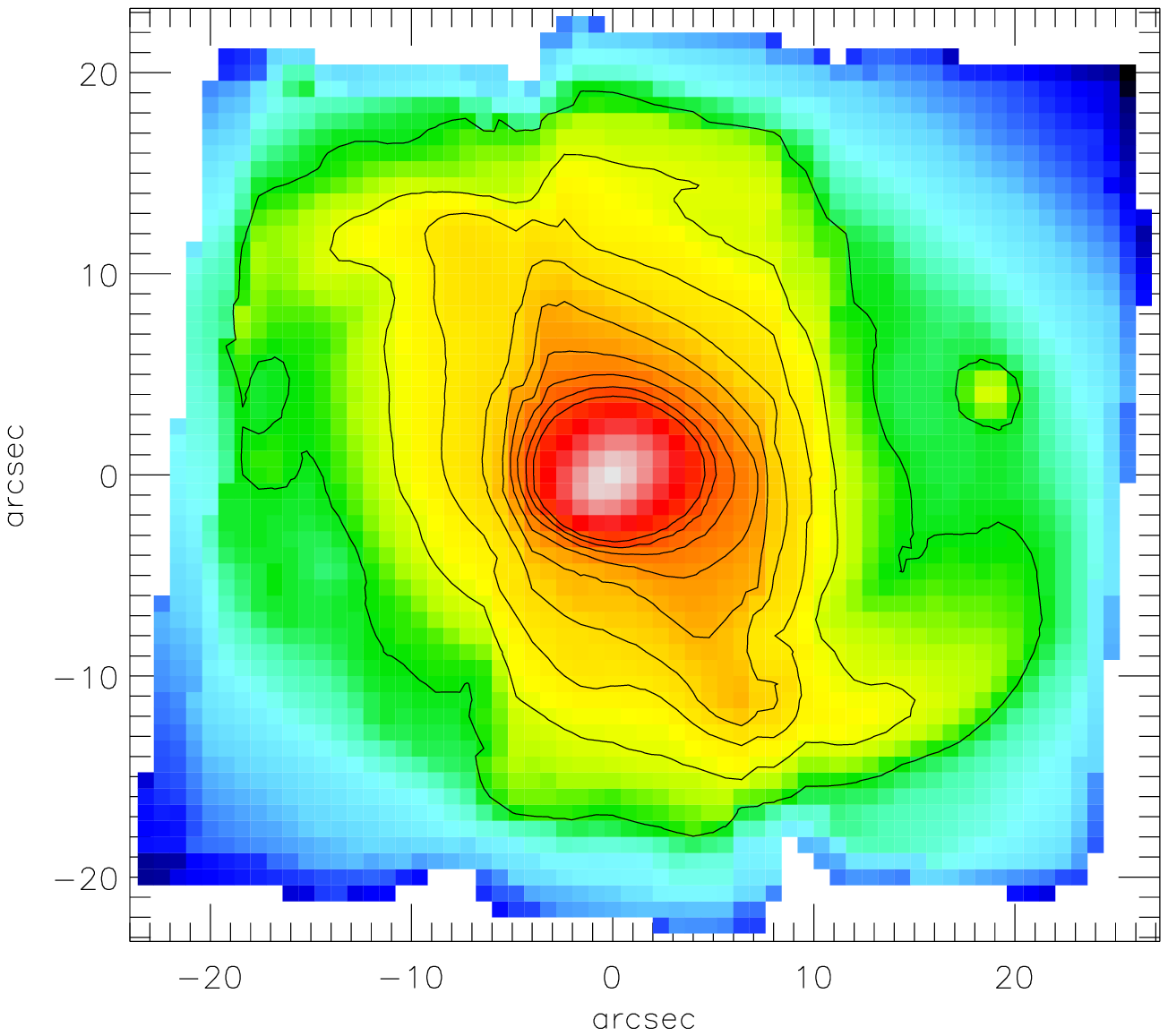}
    \caption{\sauron{} observation of the barred Sa galaxy NGC~2273. \emph{Top panel:} reconstructed total intensity: the bar is clearly visible. Contours are superimposed in 0.5 mag arcsec$^{-2}$ steps. \emph{Bottom panel:} map of the average \SN\ per spectral element. Contours are superimposed in the range 10--50 with steps of 5. Note the two vertical \SN{} jumps close to the middle of the frame and the irregular boundaries due to the merging process of the different exposures.}
    \label{fig:n2273-SNmap}
\end{figure}

In what follows we consider different methods and we apply each one to
observations of the barred Sa galaxy NGC~2273
(\reffig{fig:n2273-SNmap}) taken with the panoramic IFS \sauron{}
\citep{sauron-paper1,sauron-paper2} at the 4.2-m William Herschel
telescope on La Palma. These observations, carried on in March 2001,
are based on two pointings of $4\times 1800$~s exposures. They cover a
field-of-view of $49'' \times 44''$ with an effective spatial sampling
of $0\farcs8 \times 0\farcs8$, over the 4810--5350~\AA{} spectral
range.  They have been selected for having high \SN{} contrast between
the inner and the outer parts (in the range $\sim 1$--$150$) and a
complex \SN-distribution, caused by the presence of spiral arms and
the merging of multiple exposures (leading to \SN-jumps and irregular
outer boundaries, see \reffig{fig:n2273-SNmap}). In the following
experiments our goal is to bin these data with a target \SN{} of
$(\SN)_{T} = 50$ per spectral element on average.

\subsection{Optimal weighting of pixels}
\label{sec:optim_weigth}

Binning is not magic: in 2D, as in 1D, some useful information has to
be present in the pixels for any binning method to be effective. In 1D
one usually restricts the spatial region to bin, along the slit, to
some range where the pixels are not fully dominated by instrumental
noise. The same can be done in 2D by discarding pixels that contain
virtually no signal from the object under examination. This is not
necessary for the \sauron{} observations discussed here, where the
noise in all the pixels is essentially Poissonian ($\mathcal{N}_i
\approx \sqrt{\mathcal{S}_i}$).

An alternative approach is to optimally weight the pixels during the
summation in a bin: for a given set of pixels, the weights for which
the sum (\ref{eq:sn_sum}) provides the maximum \SN{} are given by
$w_i=\mathcal{S}_i/\mathcal{N}^2_i$ \citep[e.g.][]{rob86}. This
weighting makes full use of photon-noise dominated pixels ($w_{i}
\approx 1$), and automatically eliminates from the summation pixels
that contain virtually no signal ($w_{i}\approx0$).
Equation~(\ref{eq:sn_sum}) then becomes:
\begin{equation}
    (\SN)_{1+2} = \sqrt{(\SN)_1^2+(\SN)_2^2}.
    \label{eq:sn_optimal}
\end{equation}

All the following considerations apply unaltered irrespective of
whether equation~(\ref{eq:sn_optimal}) is used instead of
equation~(\ref{eq:sn_sum}) for the estimation of the \SN{} of each
bin. Moreover the two equations automatically coincide in the limit of
Poissonian noise and in that case no distinction needs to be made.
However, the price to pay for the use of the optimal weighting is that
the weights of each pixel contributing to a bin have to be recorded
for a complete description of the bin. This complicates the
quantitative interpretation of binned data (e.g., in dynamical
modeling).

\section{Quadtree method}
\label{sec:quadtree}

It is instructive to first consider the Quadtree algorithm
\citep{Samet84}, which we believe is close to the best `regular' image
processing method that can be used for the present application. We
show that the Quadtree method cannot produce an \emph{optimal} binning
--- that is satisfying all three previous requirements --- so that
more complex schemes are required. These will be presented in
\refsec{sec:vt}.

The Quadtree method consists of a recursive partition of a region of the
plane into axis-aligned squares. Initially one square, the \emph{root},
covers the entire region. Subsequently each square, whose \SN\ is higher
than a given threshold, is divided into four equal \emph{child} squares, by
splitting with horizontal and vertical segments through its center. The
collection of squares then form a tree, with the \emph{root} square at the
top and with smaller squares at lower levels of the tree.

\begin{figure}
  \plottwo{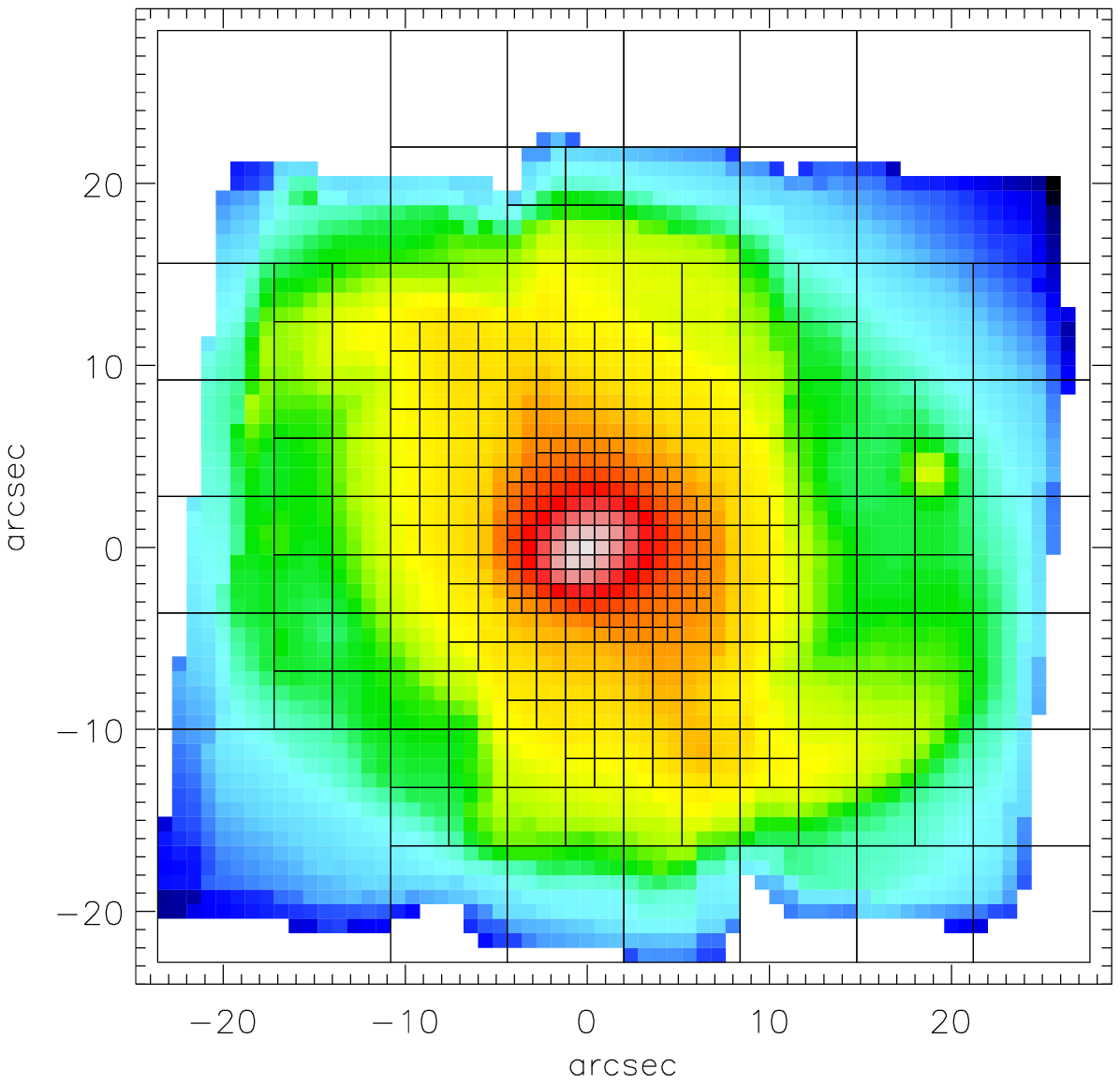}{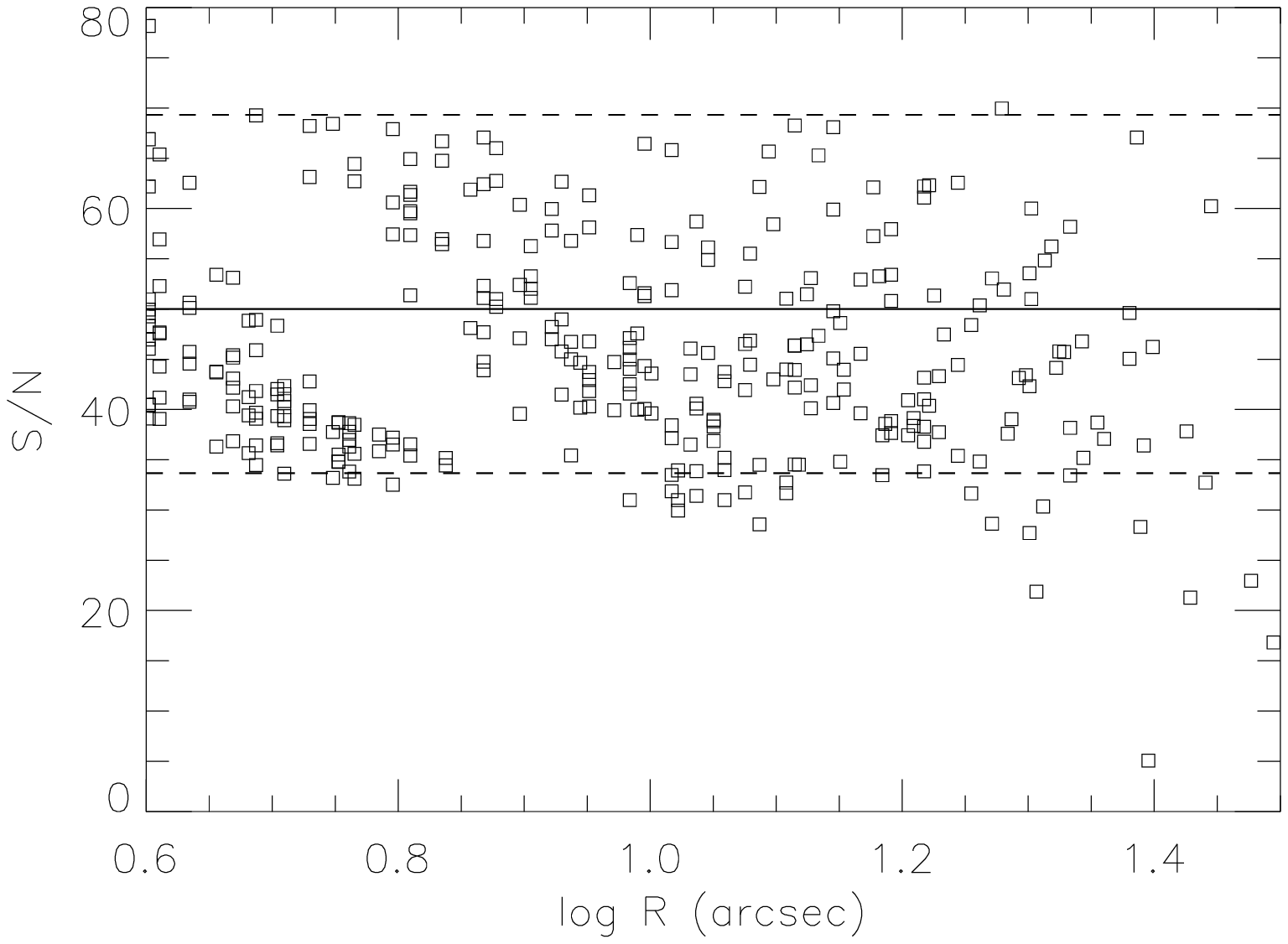}
    \caption{\emph{Top panel:} Quadtree binning of NGC~2273 for a
      target $(\SN)_T=50$. The bins are overlaid on the S/N map.
      \emph{Bottom panel:} \SN{} scatter in the above binning around
      $(\SN)_T$ (solid line): the squares represent the \SN{} of each
      bin, as a function of its distance from the galaxy center. The
      central regions ($R<4''$), where the pixels are already above
      $(\SN)_T$ and where binning is not needed, are not shown in this
      plot, as in the following ones. The dashed lines represent the
      limits of the natural scatter, that is
      $(\SN)_T\times(\sqrt{2})^{\pm 1}$.}
    \label{fig:quadtree_scatter}
\end{figure}

In \reffig{fig:quadtree_scatter} the Quadtree method was used to rebin
the \SN{} map of \reffig{fig:n2273-SNmap} into squares satisfying a
minimum \SN{} requirement. The nice feature of this binning method is
that the resulting bins are squares of various sizes (except at the
border). In this way bins are easy to handle, and require little
information to be described completely.

There are however two major shortcomings with this method:
\begin{enumerate}
\item a \SN{} spread of a factor $\sim2$ is unavoidable due to the
  fact that the bin area varies by construction in steps of a factor
  of~4;

\item unless the original image has a linear size, in pixels, which is
  a power of two, some bins at the border will not be square and
  generally will not meet the minimum \SN{} criterion, becoming
  unusable for later analysis.
\end{enumerate}

\section{Voronoi tessellation}
\label{sec:vt}

Given the inability of methods employing `regular' bins to produce
optimal 2D tessellations, we now consider schemes which do not have
square or rectangular bins. Accordingly we consider the \emph{Voronoi
  Tessellation} (VT) that can be used to generate binnings satisfying
all the three requirements of \refsec{sec:generic}.

Given a region $\Omega$ of the plane and a set of points
$\{\mathbf{z}_{i}\}_{i=1}^N$ of $\Omega$, called \emph{generators}, a
VT is a partition of $\Omega$ into regions $\{V_{i}\}_{i=1}^N$
enclosing the points closer to $\mathbf{z}_{i}$ than to any other
generator. Each $V_{i}$ is referred to as the \emph{Voronoi region}
and here as \emph{bin} associated with $\mathbf{z}_{i}$ \citep[see
e.g.][for a comprehensive treatment]{okabe00}.

The VT presents many interesting features for the binning problem:
\begin{enumerate}
\item it naturally enforces the Topological requirement;

\item it is efficiently described by the sole coordinates of its
  generators;

\item it is very easy to implement in the discrete case: given the
  generator positions, it is sufficient to locate the closest
  generator to any given pixel to determine the bin to which it
  belongs.
\end{enumerate}

\begin{figure}
  \plottwo{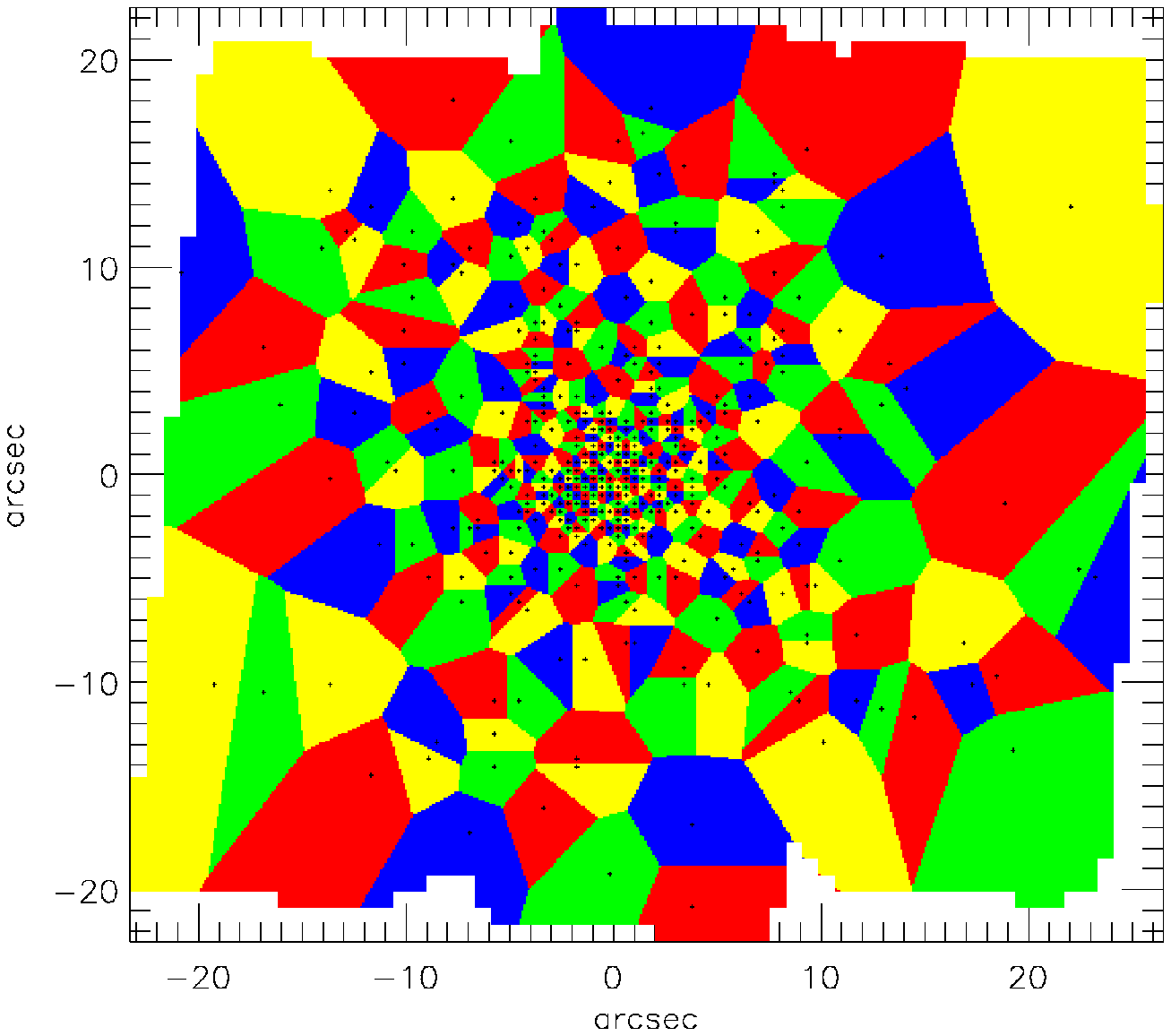}{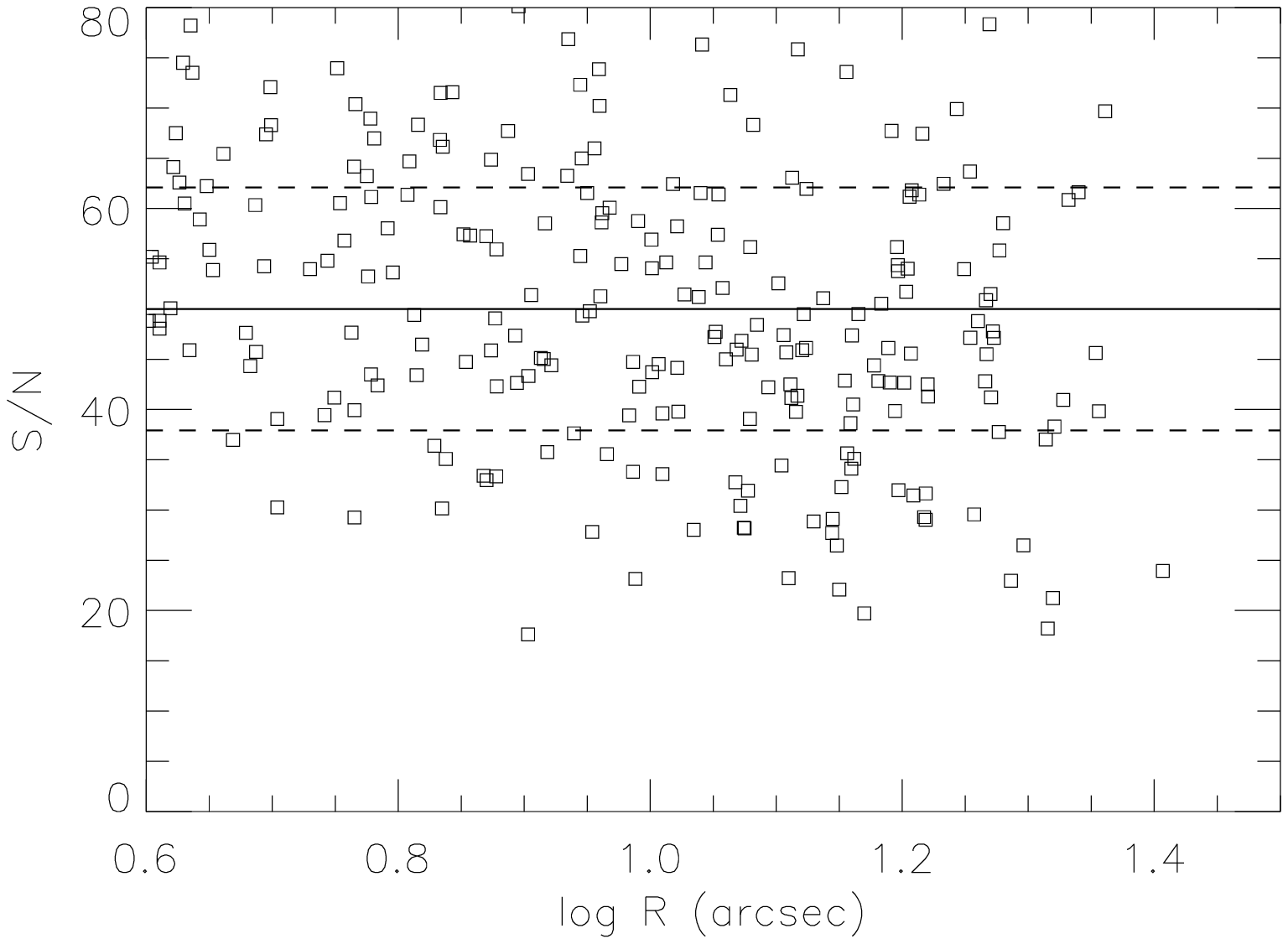}
    \caption{\emph{Top panel:} Monte Carlo VT of NGC~2273. The
      probability for a generator to be at location $\mathbf{r}$ is
      proportional to $\rho(\mathbf{r})$, where the local density
      $\rho$ is proportional to the $(\SN)^2$ in
      \reffig{fig:n2273-SNmap}. This ensures that the \SN\ enclosed in
      each bin is constant on average. The generators of the VT are
      indicated by the black dots. \emph{Bottom panel:} \SN{} scatter
      in the above VT. Although the values fluctuate around the
      desired $(\SN)_{T} = 50$ (solid line), the scatter is large: RMS
      error is 26\% (dashed lines).}
    \label{fig:Monte-Carlo}
\end{figure}

On the other hand, the fact that a VT is adopted for binning does not
enforce by itself the Morphological requirement: the bins are convex by
construction, but can have very sharp angles. Furthermore, the Uniformity
requirement is not addressed in any way by the simple use of a VT. These
two shortcomings are clearly illustrated\footnote{The four-coloring of
\reffig{fig:Monte-Carlo}, \ref{fig:n2273-cvt} and \ref{fig:bubbles-final},
was done by first computing the Delaunay triangulation (see Appendix~A)
from the VT generators and then optimally coloring the corresponding graph
with the \emph{Mathematica} package Combinatorica by \citet{pem03}.} in
\reffig{fig:Monte-Carlo}.  The Morphological and Uniformity requirements
have to be tackled through a properly tailored distribution of the Voronoi
generators. We present now a way to produce such a distribution.

\subsection{Centroidal Voronoi tessellation}
\label{sec:cvt}

In the two important cases where the noise is Poissonian or when the
pixels in each bin are optimally weighted (\refsec{sec:optim_weigth}),
the $(\SN)^2_\mathrm{bin}$ of one bin is computed by simply summing
the $(\SN)^2_i$ of the corresponding pixels
(equation~[\ref{eq:sn_optimal}]). One can then define a density
distribution $\rho(\mathbf{r})=(\SN)^2(\mathbf{r})$ such that the
problem of binning to a constant \SN{} reduces to that of obtaining a
tessellation enclosing equal mass according to $\rho$. The
\emph{Centroidal Voronoi Tessellation} (CVT) is a technique which can
be used to generate this optimally regular \emph{and} uniform VT in
the continuous case, or in the limit of a large number of pixels.

Given a density function $\rho(\mathbf{r})$ defined in a region
$\Omega$, a CVT of $\Omega$ is a special class of VT where the
generators $\mathbf{z}_{i}$ happen to coincide with the mass centroids
\begin{equation}
  \mathbf{z}_{i}^{\ast} =
  \frac{\int_{V_{i}} \mathbf{r}\rho(\mathbf{r})\,\d\mathbf{r}}
  {\int_{V_{i}} \rho(\mathbf{r})\,\d\mathbf{r}}
\end{equation}
of the corresponding Voronoi regions $V_{i}$. As illustrated in the
review by \citet*{du1999-CVT}, the CVTs are useful to solve a variety
of mathematical problems, but can also be observed in many real-life
natural examples (living cells, territories of animals, etc.).

One of the most striking characteristics of CVT in the 2D-case is its
ability to partition a region into bins whose size varies as a
function of the underlying density distribution, but whose shape tends
asymptotically to a hexagonal-like lattice for a large number of bins.
Another nice feature of CVT is that a simple algorithm exists for its
practical computation: the \cite{lloyd82} method, for which the CVT is
a fixed point.

Although CVT bins are naturally smaller where the density is higher,
the $\mathrm{area}-\rho$ relation of the bins is not such that the
mass enclosed in every bin is constant: the CVT cannot be used
directly to produce equal mass bins (equal \SN{} in the case of photon
noise or with optimal pixels weighting, see \refsec{sec:optim_weigth}).
However, we now present a modification of the Lloyd algorithm allowing
it to converge toward an equi-mass 2D-CVT.

\paragraph*{1D-case.}
\label{sec:1d-case}

It has been demonstrated that in 1D, and asymptotically for large
numbers of bins (which in 1D are intervals), the size $d$ of the CVT
bins is proportional to the one-third power of the underlying density
at the midpoint of the bin \citep{du1999-CVT}: $d \propto
\rho^{-1/3}$. Since the mass enclosed in each bin is $m \approx
\rho\,d$, one has $m \propto \rho^{2/3}$: the mass in the CVT bins is
only constant for a constant density.

However a CVT can still be used to partition a segment into equal mass
intervals. In fact, one can derive from the previous relations that,
if the CVT is performed on a different density $\rho' = \rho^{3}$, the
partition found induces equal-mass bins on the original density
$\rho$: $m \approx \rho\,d \propto \rho\,\rho'^{-1/3} = 1$.  This
result is not being used in 1D, but can be generalized in 2D (or
higher dimensions) to produce bins that have the nice regularity
properties of the CVTs but, in addition, enclose equal mass.

\paragraph*{2D-case.}
\label{sec:2d-case}

\begin{figure}
    \plotone{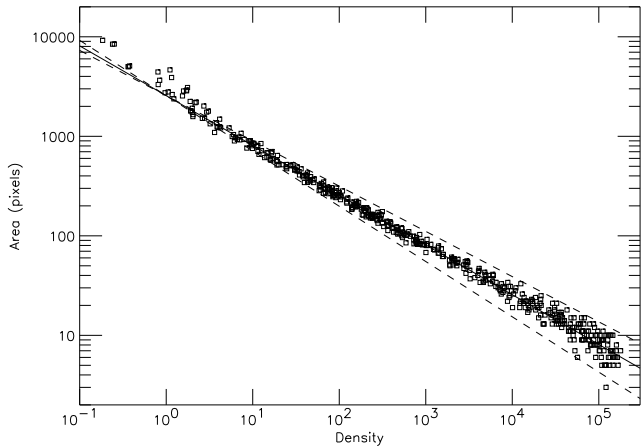}
    \caption{The $\mathcal{A}-\rho$ relation of a two-dimensional
      CVT. Over-plotted is the best fitting relation $\mathcal{A}
      \propto \rho^{-1/2}$ (solid line) and, for comparison, the two
      relations $\mathcal{A} \propto \rho^{-1/2.2}$ and $\mathcal{A}
      \propto \rho^{-1/1.8}$ (dashed lines). The relation is a perfect
      straight line, with the exception of the largest bins, closer to
      the border, and the smallest ones, far from the continuum
      approximation.}
    \label{fig:area-density}
\end{figure}

In 2D, the so-called Gersho conjecture \citep{gersho79,gray98} can be
used to predict the corresponding relation between the area
$\mathcal{A}$ of a CVT bin and the underlying density $\rho$.  A
consequence of this conjecture is the \emph{principle of equal
  partition of error}. In 2D, it implies that, if $d$ is the typical
size of the CVT bin, $\rho\,d^{4}$ is roughly constant\footnote{More
  generally, in $n$D, one has $d \propto \rho^{\frac{-1}{2+n}}$.}.
Since $\mathcal{A} \propto d^{2}$, one gets the relation $\mathcal{A}
\propto \rho^{-1/2}$. Although a rigorous proof of this relation is
not known (Du, private communication), we have verified with various
numerical experiments that this relation is indeed very accurately
verified in practice, even with a small number of bins
(\reffig{fig:area-density}).

Applying the same reasoning in 2D as in the 1D-case, to generate
equal-mass bins one has to satisfy the relation $m \approx
\rho\,\mathcal{A} \propto \rho\, \rho'^{-1/2} = 1$, which implies
$\rho' = \rho^{2}$. We have thus shown that \emph{a CVT of the density
  distribution $\rho' = \rho^{2}$ generates a VT with bins that
  asymptotically enclose a constant mass according to the density
  $\rho$}.

Given a region $\Omega$, a density function $\rho$ and $N$ generators
$\{\mathbf{z}_{i}\}_{i=1}^N$, our modified Lloyd method to produce an
equi-mass 2D-CVT is thus the following:
\begin{enumerate}
\item select an initial random set $\{\mathbf{z}_{i}\}_{i=1}^N$ of
  positions for the generators: the probability for a generator to be
  at position $\mathbf{r}$ is proportional to $\rho(\mathbf{r})$;

\item \label{item:lloyd2} perform a VT of $\Omega$ associated with the
  generators $\{\mathbf{z}_{i}\}_{i=1}^N$: with the above initial
  generator distribution, the Voronoi regions $\{V_{i}\}_{i=1}^N$
  contain \emph{on average} a constant mass, but the scatter is large
  and the bins are generally badly shaped (as in
  \reffig{fig:Monte-Carlo});

\item compute the mass centroids of the $\{V_{i}\}_{i=1}^N$ according
  to the density $\rho' = \rho^{2}$: these constitute now the new set
  $\{\mathbf{z}_{i}\}_{i=1}^N$ of generators\footnote{The original
    Lloyd method uses $\rho' = \rho$ and thus does not enforces equal
    mass bins.};

\item iterate over step \ref{item:lloyd2} until the coordinates of the
  VT generators don't change any more.
\end{enumerate}

\begin{figure}
  \plottwo{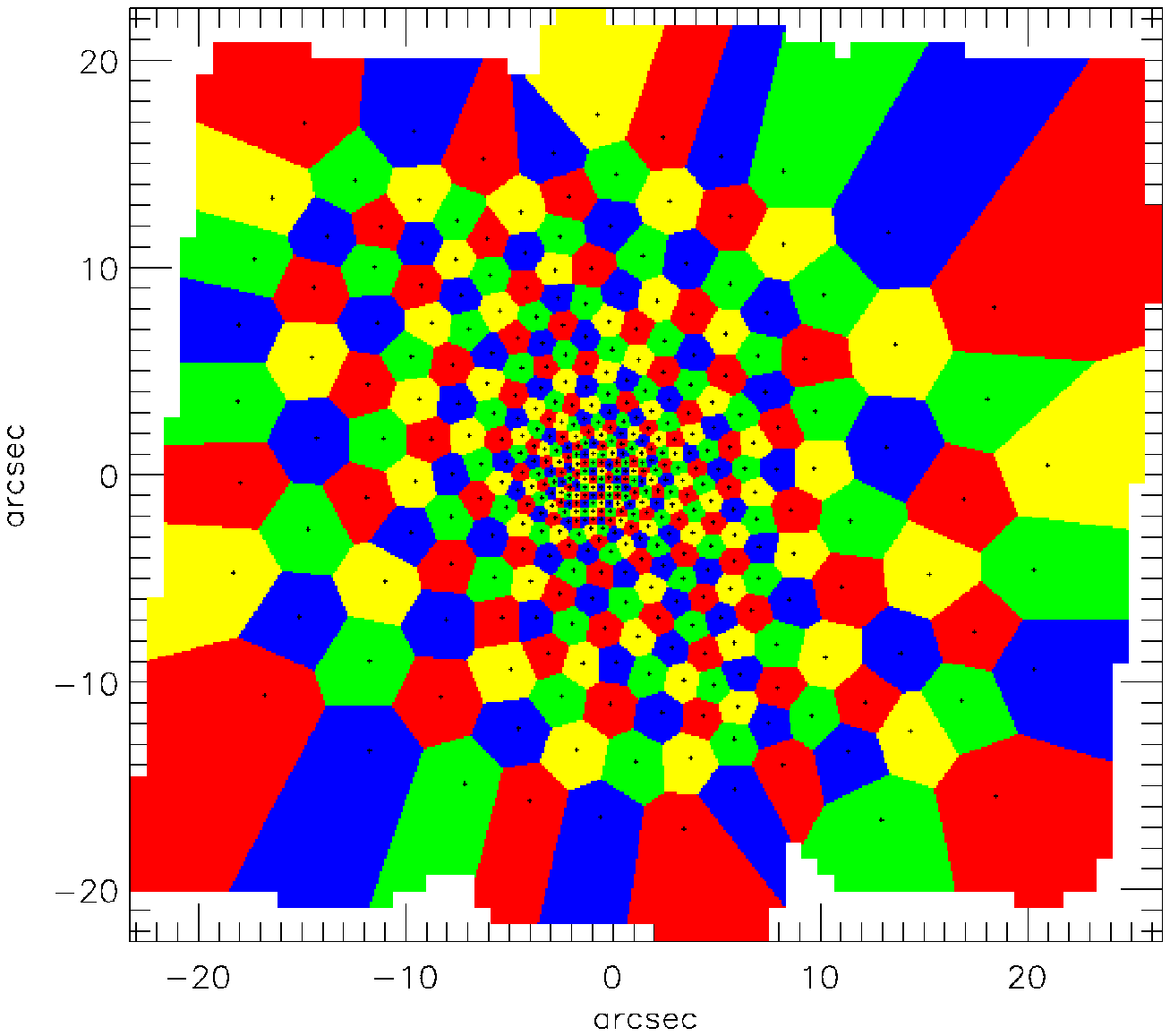}{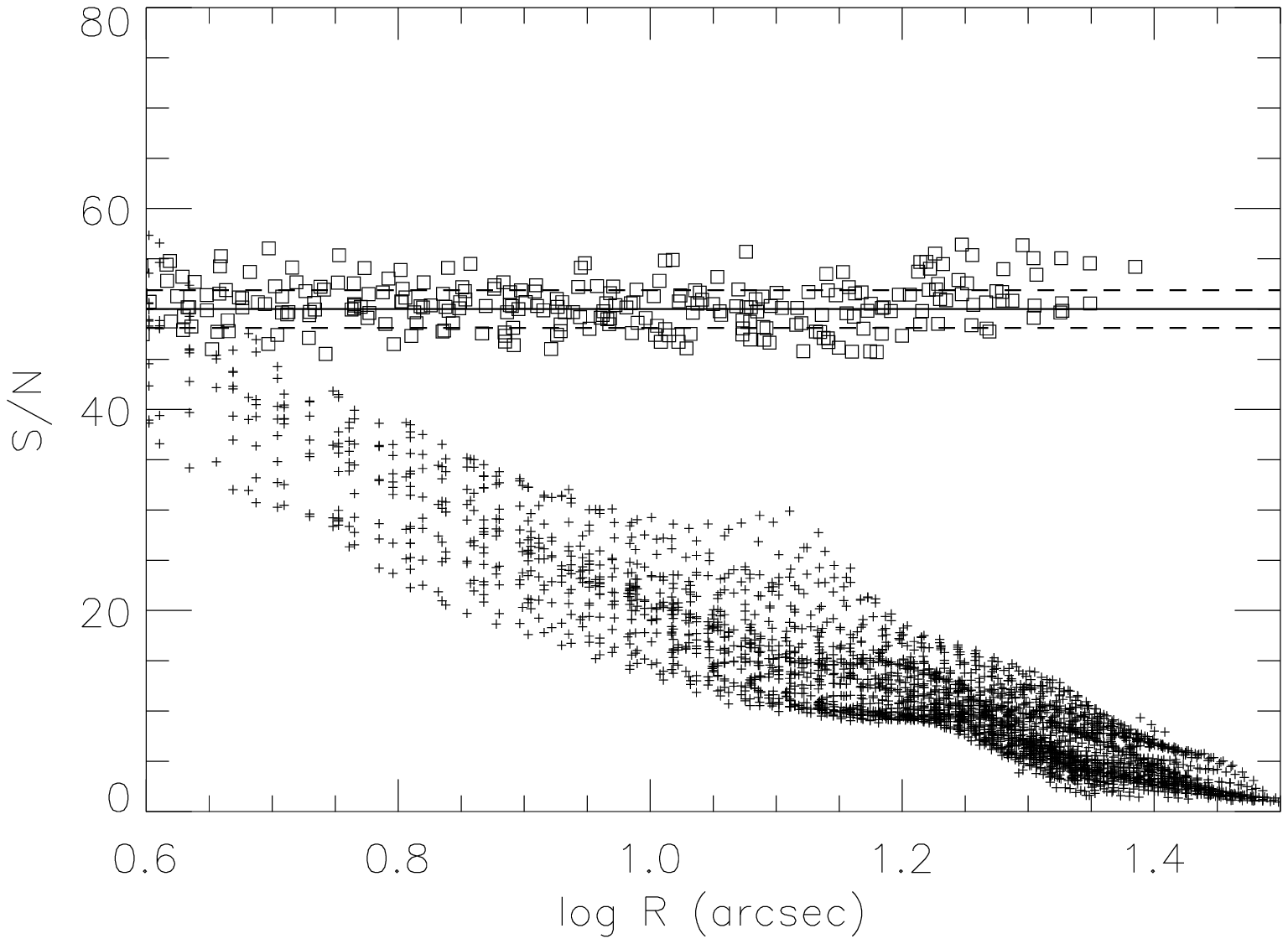}
    \caption{Continuous case. \emph{Top panel:} CVT-binning of the
      `continuous' density $\rho'(\mathbf{r})=\rho^{2}(\mathbf{r})$,
      where $\rho$ is obtained by linear interpolation from the
      $(\SN)^2$ in \reffig{fig:n2273-SNmap}. The generators of the CVT
      are indicated by the black dots. \emph{Bottom panel:} \SN{}
      scatter in the above CVT. The \SN{} of the original pixels is
      shown with the crosses, while the squares represent the \SN{} of
      the final bins. The target $(\SN)_{T} =50$ is indicated by the
      solid line. The RMS scatter is $\sim3$\% (dashed lines).}
    \label{fig:n2273-cvt}
\end{figure}

\reffig{fig:n2273-cvt} presents a CVT produced by applying the
previous algorithm to the density $\rho$, obtained by linearly
interpolating the $(\SN)^2$ of NGC~2273 in \reffig{fig:n2273-SNmap}
onto a grid with pixel size $8\times$ smaller than the original size.
This interpolation is used here to approach the continuous case and
ensure a better convergence of the modified Lloyd algorithm.  In this
case of a large number of pixels the bins of the VT tend to the
theoretical hexagonal shape and adapt nicely to density variations and
to the irregular boundary of the region. The scatter of the \SN{} is
also close to optimal with RMS scatter of $\sim 3$\%.

\begin{figure}
  \plottwo{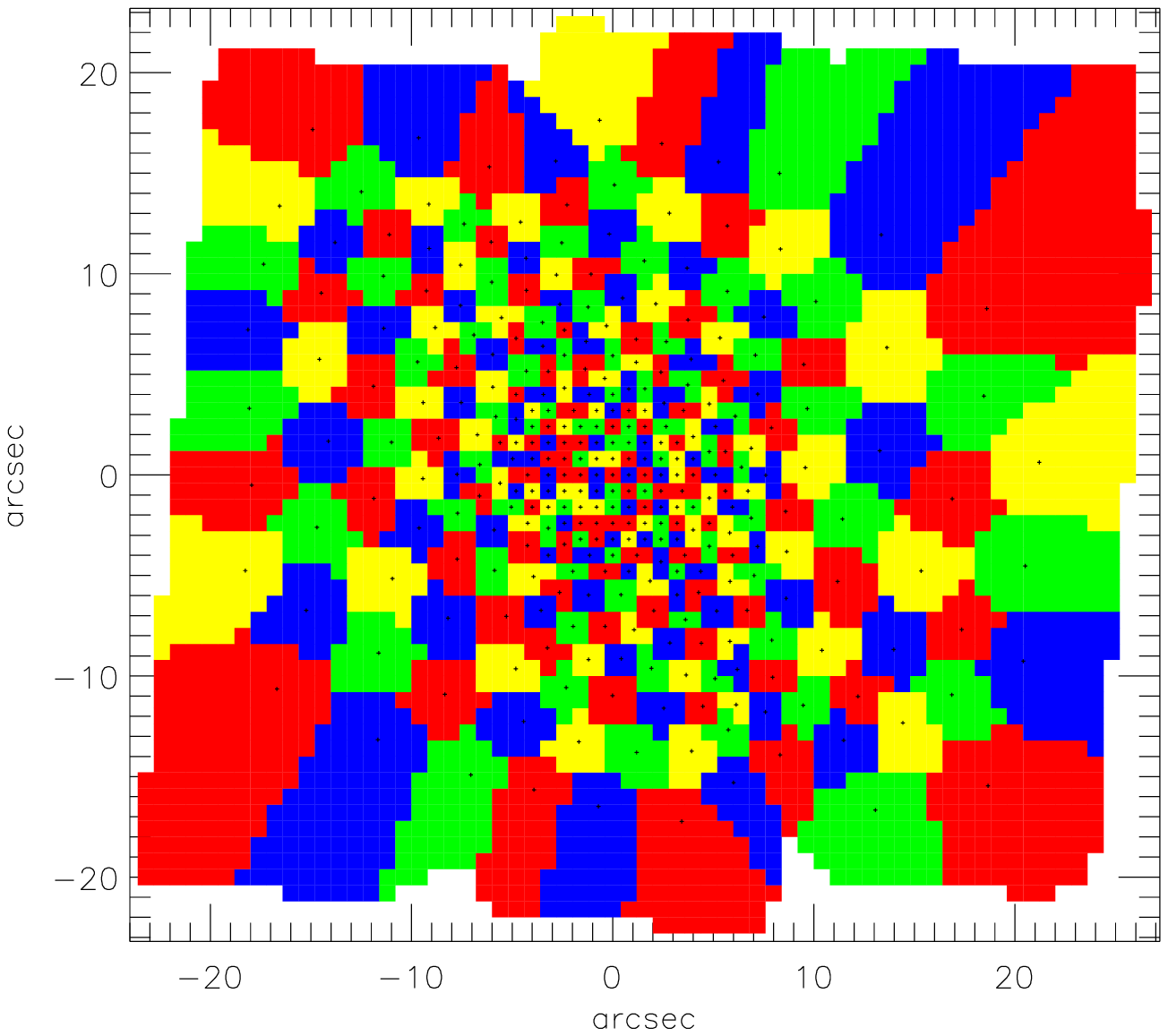}{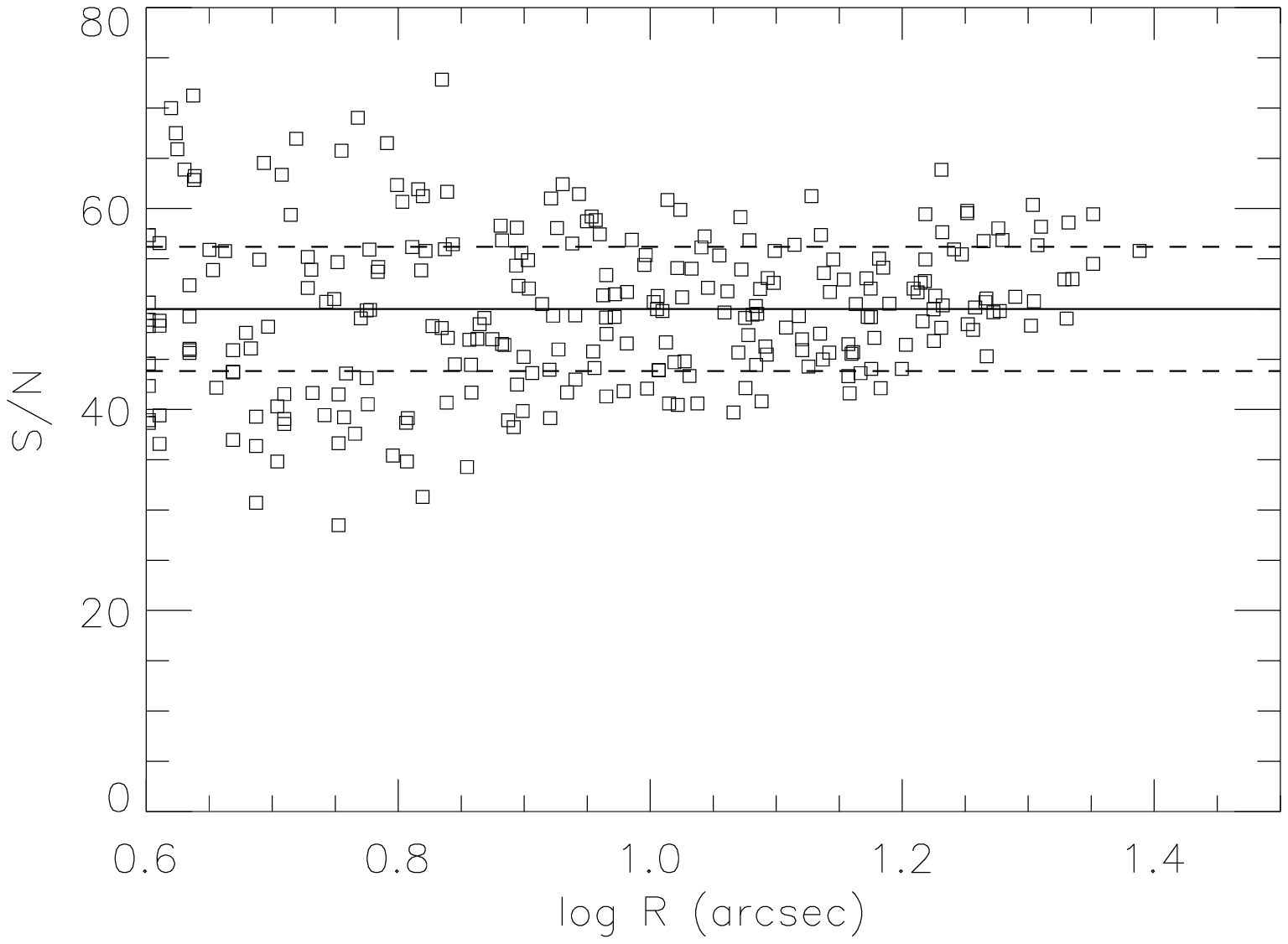}
    \caption{Discrete case. \emph{Top panel:} VT of the \sauron{}
      pixel grid of \reffig{fig:n2273-SNmap}, obtained using the same
      generators (black dots) as in \reffig{fig:n2273-cvt}.
      \emph{Bottom panel:} \SN{} scatter in the above VT, as a function
      of the distance from the galaxy center. Note the significant
      increase of the scatter (RMS of $\sim14$\%, dashed lines) around
      $(\SN)_{T}$ (solid line), compared to \reffig{fig:n2273-cvt},
      due to discretization effects.}
    \label{fig:n2273-cvt-bad}
\end{figure}

This CVT method illustrates the goals towards which an optimal
2D-binning algorithm should tend, but it has still some practical
limitations:
\begin{enumerate}
\item it generates equal-mass bins but \emph{not} necessarily
  equal-\SN{} bins, unless e.g.\ the noise is Poissonian or the bins
  are optimally weighted (see the beginning of this section).

\item more importantly the method does not work well when the bins are
  constituted of just a few pixels. This is due to the fact that the
  Lloyd method is designed to work in a continuum approximation and
  does not necessarily converge to the global minimum in a discrete
  case, unless a good set of initial generators is chosen
  \citep{du1999-CVT}.  In \reffig{fig:n2273-cvt-bad} the same
  generators as in \reffig{fig:n2273-cvt} where used to construct a VT
  for the coarser \sauron{} pixel grid of \reffig{fig:n2273-SNmap}.
  The obtained VT is similar to that of \reffig{fig:n2273-cvt}, but
  the \SN{} scatter increases considerably, due to discretization
  effects. Constructing the bins from an oversampled and interpolated
  grid is not an acceptable solution if an accurate solution is
  desired, as this would introduce correlations within the bins. One
  wants to construct the bins by coadding actually observed pixels.
\end{enumerate}

In practice an optimal 2D-binning method has to preserve the good
characteristics of the equi-mass 2D-CVT, in the limit of many pixels,
but has to be able to take the discrete nature of pixels into account,
when dealing with bins constituted by just a few pixels. This
algorithm is the subject of the next section.

\subsection{Bin-accretion algorithm}
\label{sec:accretion}

Finding a good set of initial generators is crucial for the Lloyd
algorithm to converge to the global minimum, whenever the discrete
nature of pixels is important. In this section we describe a method we
developed to find the generators for the optimal VT taking the
discrete nature of pixels into account from the beginning. The
algorithm described here constructs an initial binning trying to
generalize to 2D the standard pixel-by-pixel binning algorithm used in
1D. The centroids of the bins found in this way are then used as
starting generators for a CVT. The method reduces to the previous CVT
in the limit of many pixels, but works on a pixel basis with bins made
by just a few pixels.

A natural 1D-binning algorithm proceeds as follows: start a bin from
the highest \SN{} unbinned pixel and accrete neighboring pixels; when
the target \SN{} is reached a new bin is started and this process is
repeated until all pixels have been binned. To extend this idea to 2D,
we need to make a clever choice regarding the pixel to be accreted
next, so that the topological and morphological constraints are
naturally enforced. The adopted method always tries to add to the
current bin the pixel that is closest to the bin centroid.
Furthermore, when a new bin is started, the first pixel is also
selected as the one closest to the centroid of all the previously
binned pixels. This simple scheme automatically \emph{tends to}
generate bins that are compact, and the bin \SN{} can be carefully
monitored on a pixel-by-pixel basis during the accretion phase. The
centroids of the bins found in this way can be used as starting
generators for a CVT.

The idea described above is quite simple, but some practical problems
have to be solved for a robust implementation of the algorithm. During
the accretion of new bins around the bin centroid, the topological and
morphological criteria \emph{tend to} be satisfied by construction,
but there can be situations where this is not the case: a common
example is when the accretion process hits the boundary of the
dataset. In this situation, as new pixels are accreted, the shape of
the growing bin will be dictated more by the shape of the border of
the dataset than by the centroid criterion. As a result some bins can
become very elongated or even disconnected before the target \SN{} is
reached. If this is not prevented, the centroids of the final bins
will not always fall close to the bin center and can even lie outside
the bin. In this case the bin centroids will not provide a good set of
initial generators for the CVT and the modified Lloyd algorithm may
not converge to an acceptable solution (e.g. some bins may not reach
$(\SN)_{T}$).

To solve this situation the simple idea of the bin-accretion algorithm
has to be complicated a little by making sure that the topological and
morphological criteria are enforced for \emph{all} bins, and not only
for most of them. In practice the shape of the bin is monitored during
the accretion phase and, if the bin does not satisfy some minimal
requirements, a new bin is initiated even if the previous bin did not
meet the minimum \SN{} criterion. The bins that do not have enough
\SN{} at the end of the accretion stage will be reassigned to the
closest good bin, before computing the centroids which will be used as
starting points of the final CVT.

\section{Optimal Voronoi 2D-binning}
\label{sec:optimal}

\subsection{Algorithm}

In this section we translate the ideas illustrated in \refsec{sec:vt}
into a practical and robust algorithm that can be used to bin actual
2D data. Our optimal Voronoi 2D-binning algorithm is described in
detail by the following steps:
\begin{enumerate}
\item \label{item:baa1} start the first bin from the highest \SN{}
  pixel of the image;

\item \label{item:baa2} evaluate the mass centroid of the current bin
  and select the unbinned pixel closest to the centroid as candidate
  for addition to the bin;

\item \label{item:baa3} check whether all the following conditions are
  satisfied in that precise order:
    \begin{enumerate}
    \item \emph{topological criterion:} the new pixel is adjacent to
      the current bin;

    \item \label{item:round} \emph{morphological criterion:} by adding
      the new pixel, the `roundness' $\mathcal{R}$ (see below) of the
      current bin would remain below a chosen threshold;

    \item \label{item:baa4} \emph{uniformity criterion:} by adding the
      new pixel, the bin \SN{} would not deviate from $(\SN)_{T}$ more
      than before the addition while the \SN{} is already higher than
      a given fraction (e.g., 80\%) of the target \SN{};
    \end{enumerate}
    If all the previous criteria are fulfilled, the accretion process
    can go on: add the candidate pixel to the current bin and go back
    to step \ref{item:baa2};

  \item the accretion process of the current bin has come to an end.
    If the uniformity criterion \ref{item:baa4} was met, mark all the
    pixels in the bin as `successfully binned', otherwise as
    `unsuccessfully binned';

  \item \label{item:baa5} evaluate the mass centroid of all the pixels
    already binned and start a new bin from the unbinned pixel closest
    to this centroid; go back to step \ref{item:baa2} until all pixels
    have been binned;

  \item compute the centroid of each successful bin and reassign the
    `unsuccessfully binned' pixels to the closest of these centroids;

  \item recompute the centroid of each bin obtained from the previous
    step: these centroids are then used as initial generators of a
    CVT, which is computed with the modified Lloyd method of
    \refsec{sec:cvt}.
\end{enumerate}

The quantity $\mathcal{R}$ used in step \ref{item:round} is any
quantity measuring the `roundness' of a bin. It can be simply defined
as:
\begin{equation}
    \mathcal{R} = \frac{r_{\max}}{r_{\mathrm{eff}}} - 1,
    \label{eq:roundness}
\end{equation}
where $r_{\max}$ is the maximum distance between the centroid of the
bin and any of the bin pixels, and $r_{\mathrm{eff}}$ is the radius of
a disk of same area as the whole bin. With this definition,
$\mathcal{R} = 0$ for a perfectly circular bin and $\mathcal{R}>0$
otherwise. We found that a roundness threshold $\mathcal{R}_{\max} =
0.3$ gives good results.

\begin{figure}
  \centering \includegraphics[height=0.89\textheight]{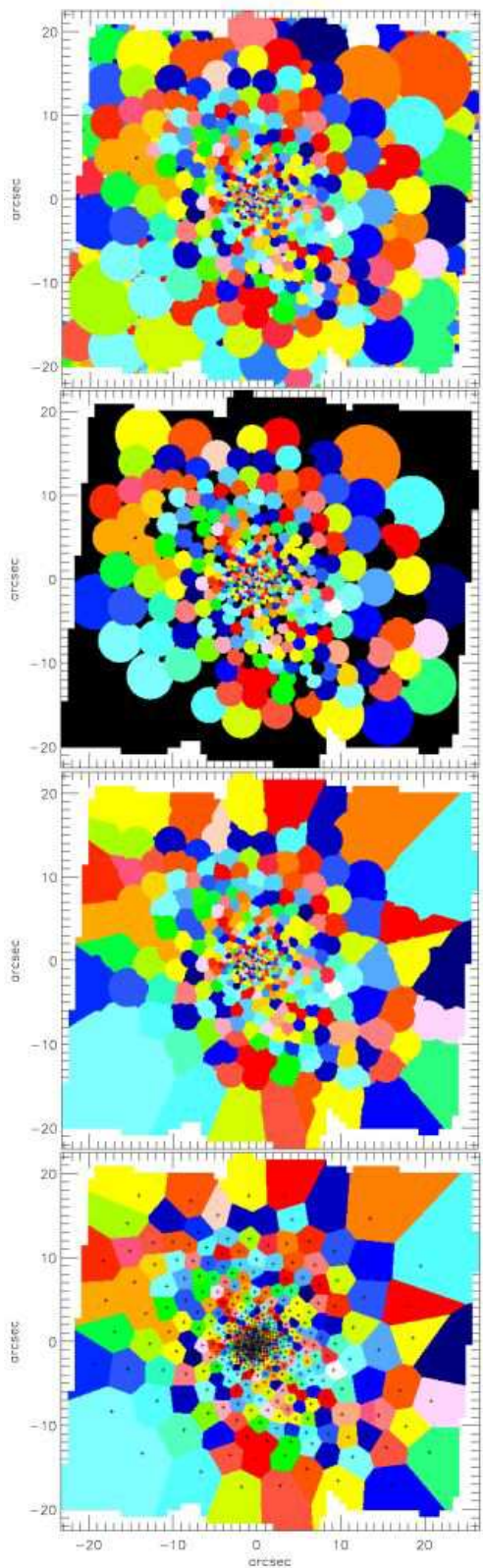}
    \caption{Four main stages of the optimal Voronoi 2D-binning algorithm
      are illustrated in the continuous case. From top to bottom: (i)
      construction of the bins by `bin-accretion', (ii) the bins that did
      not satisfy the convergence condition are flagged (black), (iii) and
      their pixels are reassigned to the closest good bin, (iv) the
      centroid of the new bins are evaluated and a VT is constructed using
      these as generators (crosses).}
    \label{fig:bin-accretion}
\end{figure}

\reffig{fig:bin-accretion} provides a visual explanation of the four
major stages of the above algorithm. For illustration purposes, it has
been applied to a `continuous' density distribution, obtained by
linearly resampling the $(\SN)^2$ of \reffig{fig:n2273-SNmap} onto a
regular $8\times$ finer grid. After a bin-accretion stage, the
unsuccessfully binned pixels are flagged and reassigned to the closest
successful bin, and a VT is generated from the centroids of the bins.

\subsection{Application to real data}

\begin{figure}
  \plottwo{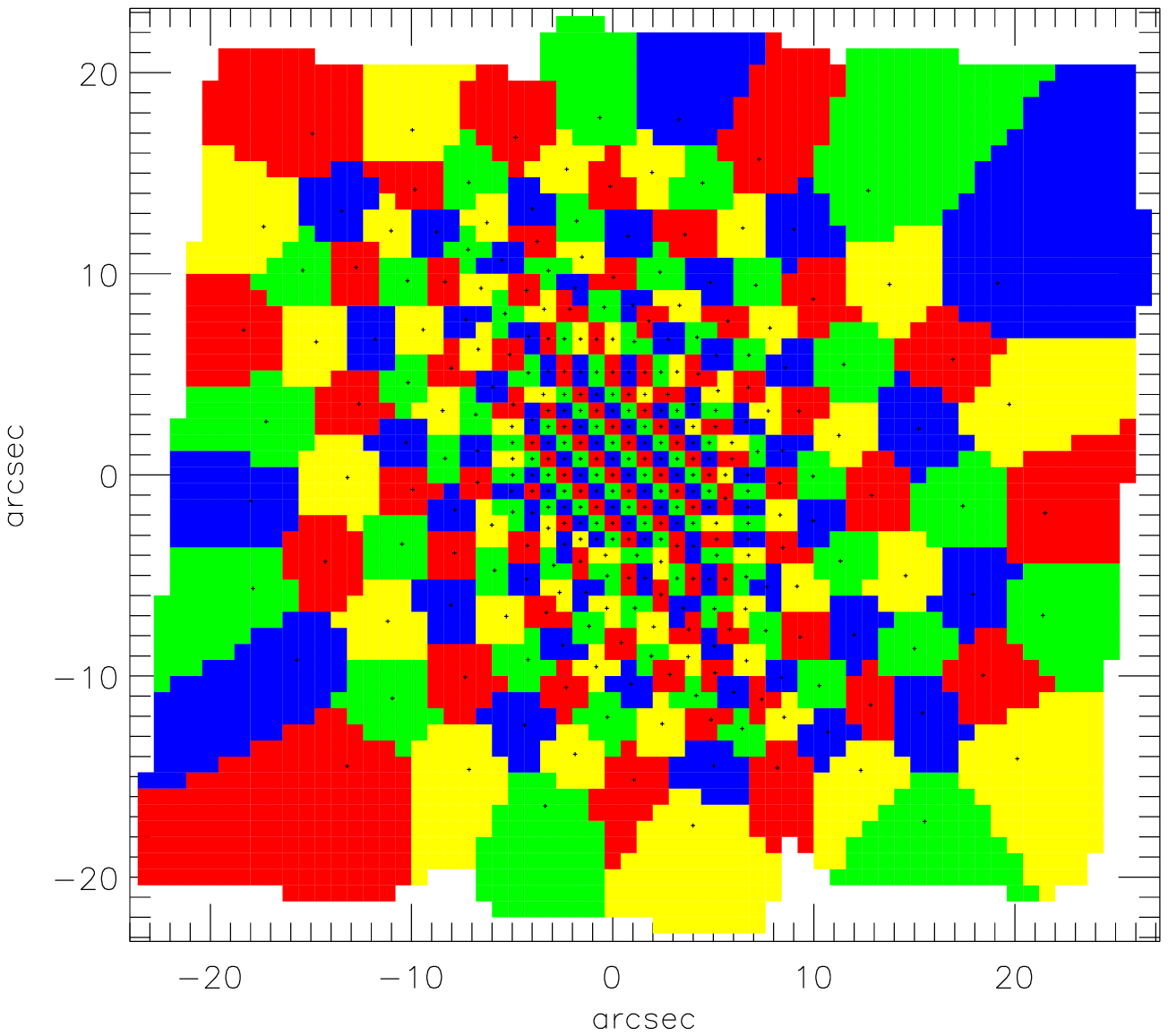}{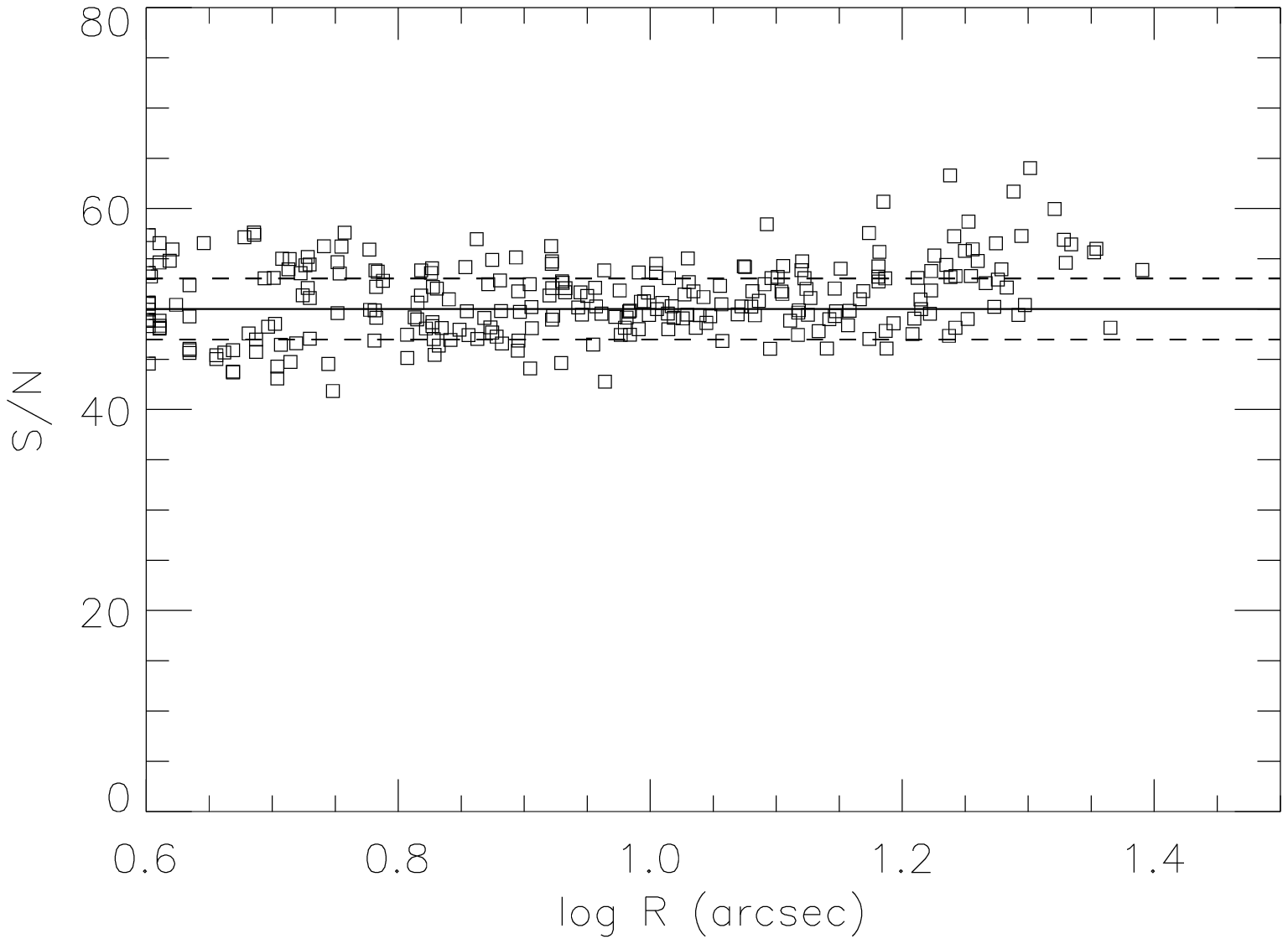}
    \caption{\emph{Top panel:} Final result for the bins after
      application of the optimal Voronoi 2D-binning algorithm to the
      \sauron{} observations of NGC~2273. \emph{Bottom panel:} \SN{} scatter
      in the above binning. The solid line represents the target \SN{}
      level, while the two dashed lines show the RMS scatter ($\sim6$\%).
      Note the decrease of the \SN{} scatter compared to
      \reffig{fig:n2273-cvt-bad}.}
    \label{fig:bubbles-final}
\end{figure}

An example of the application of our optimal Voronoi 2D-binning algorithm to
the actual \sauron{} data of NGC~2273 is shown in the top panel of
\reffig{fig:bubbles-final}, while the resulting \SN{} scatter is shown in
the bottom panel: the RMS value is $\sim 6$\%. The \SN{} values,
symmetrically clustered around the target $(\SN)_{T} = 50$, essentially
represent the lowest \SN{} scatter obtainable from a binning of these data:
all the scatter is due to discretization noise, which increases towards the
galaxy center, where the bins are made of a smaller number of pixels. This
figure shows that the method is able to guarantee a minimum \SN{} to
\emph{all} bins, while retaining a regular shape.

The binning resulting from this algorithm is qualitatively similar to
the one obtained using the CVT (see \reffig{fig:n2273-cvt-bad}). But, by
contrast to the CVT alone, this method is able to produce bins that
are essentially optimal also in the `small-bins' regime, with bins of
only 2--4~pixels.

\begin{figure}
  \plottwo{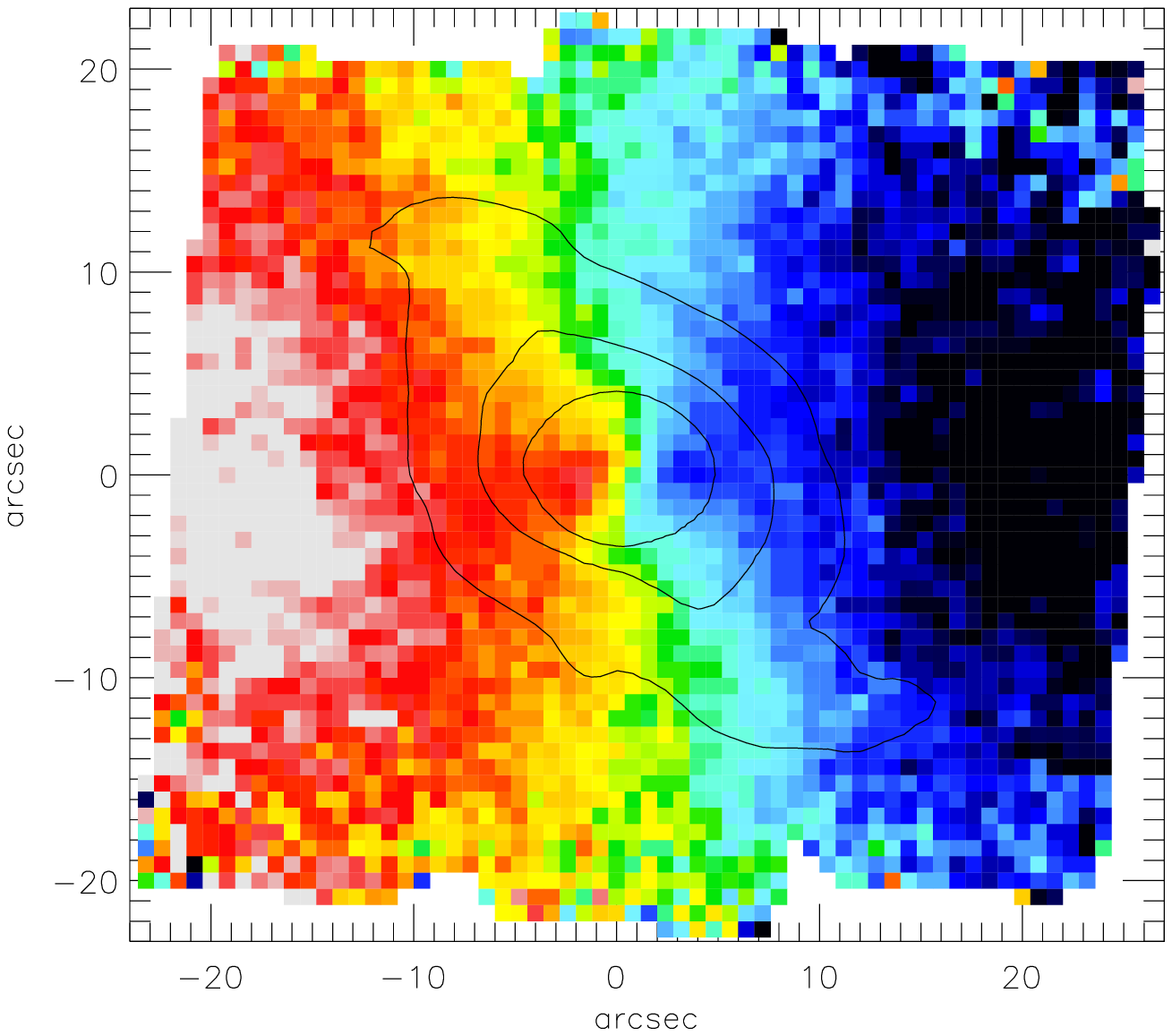}{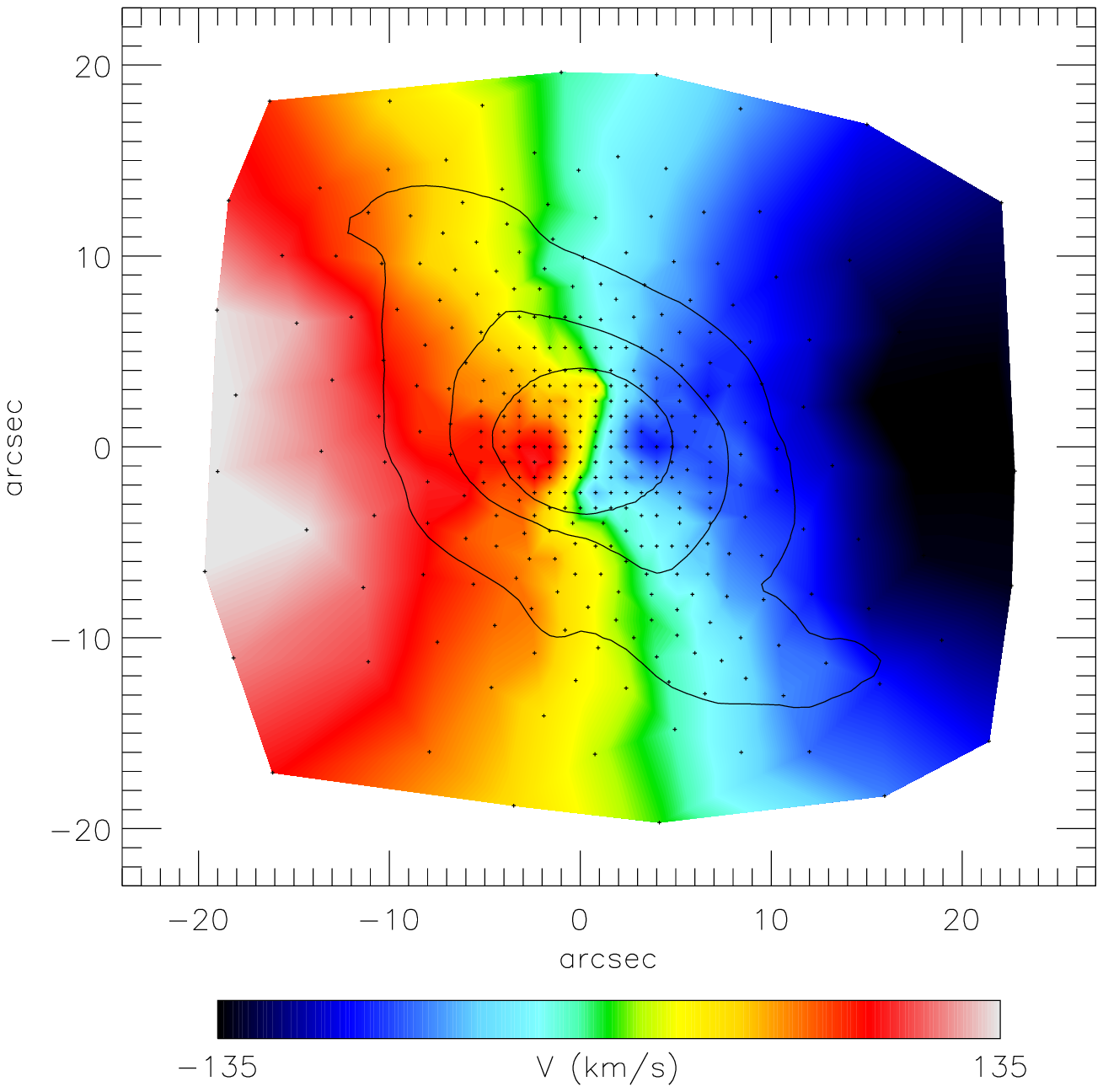}
    \caption{The stellar mean velocity $V$ field measured from the
      unbinned \sauron{} data of NGC~2273 (\emph{top panel}), is
      compared to the linearly interpolated velocity field extracted
      from the 2D-binned spectra (\emph{bottom panel}). The same color
      levels are used in both plots. The flux weighted centroids
      (\emph{not} the VT~generators) of the bins used
      (\reffig{fig:bubbles-final}) are indicated by the black dots.
      Some representative contours of the galaxy surface brightness
      are also shown. The interpolated binned velocity field
      was truncated to the centroids of the outer bins.}
    \label{fig:n2273_velfield}
\end{figure}

\begin{figure}
  \plottwo{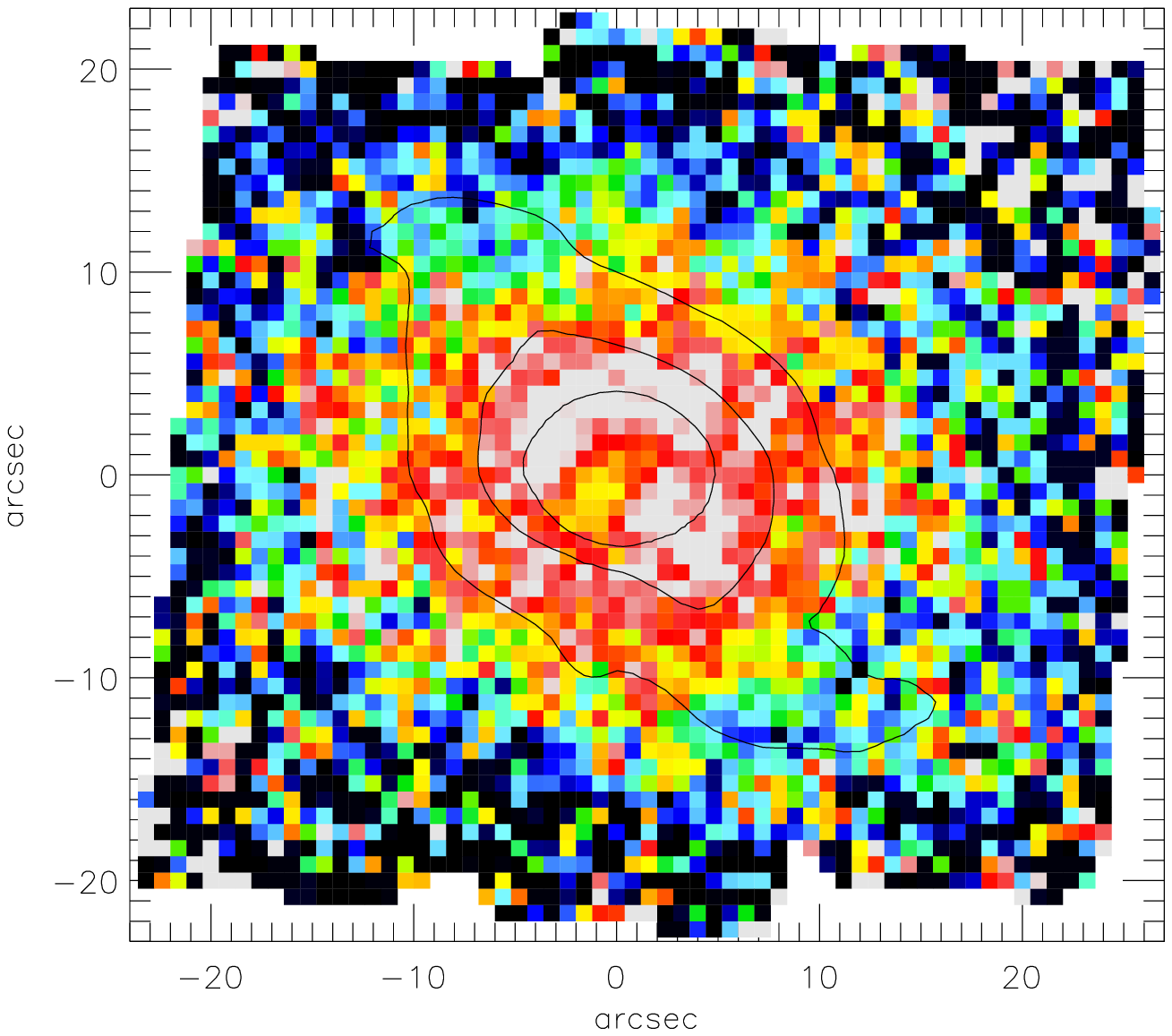}{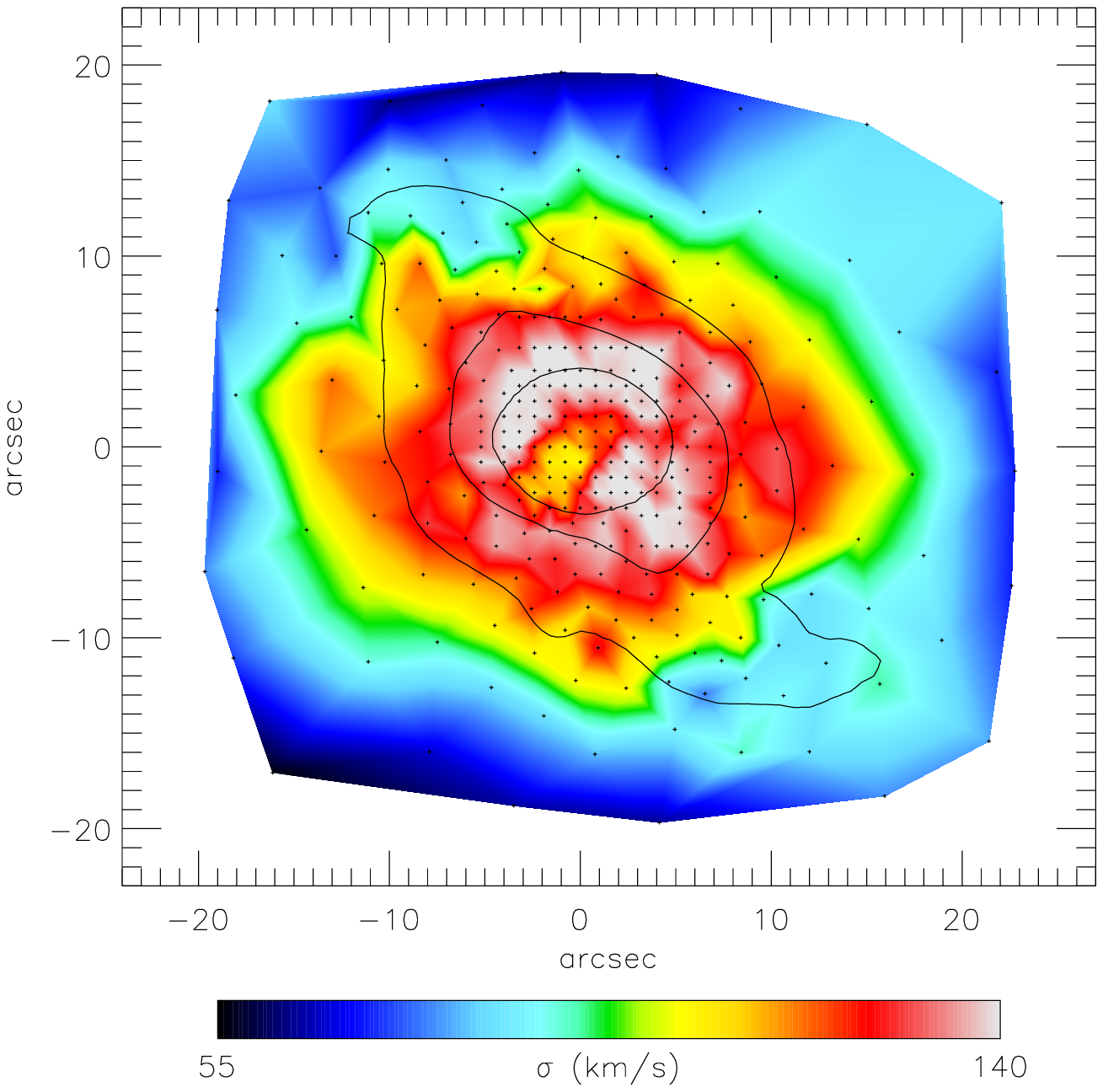}
    \caption{Same as in \reffig{fig:n2273_velfield} for the stellar
      velocity dispersion $\sigma$ field. Very little information on
      the dispersion field can be extracted without binning.}
    \label{fig:n2273_sigfield}
\end{figure}

The stellar mean velocity and velocity dispersion fields extracted
from the binned data are shown in \reffig{fig:n2273_velfield} and
\reffig{fig:n2273_sigfield}. The spectra obtained with \sauron\
include prominent gas emission lines of H$\beta$,
[\ion{O}{III}]$\lambda\lambda$4959, 5007 and [\ion{N}{I}] $\lambda$5200. The
kinematics extraction was performed in pixel-space using the method of
\citet{cap02}, and adopting a single K1~III template star observed
with \sauron. The spectral regions possibly contaminated by gas
emission lines were excluded from the fit in all spectra. Not
surprisingly, while for the unbinned data some information can be
extracted only from the galaxy central regions, where the pixels \SN{}
is naturally high enough, the use of binned data leads to consistent
measurements over the whole observed field.

The kinematics maps in \reffig{fig:n2273_velfield} and
\ref{fig:n2273_sigfield} were linearly interpolated using a
Delaunay triangulation of the measured values, which were assigned to
the position of the flux weighted centroid of each bin. If the
underlying kinematics varies smoothly over the scale of the bins, and
if the measurements errors are negligible, this approach closely
recovers the `true' kinematical field. Using an interpolation it
becomes easy to visually detect regularities in the data, like the
almost perfect point-symmetry of the velocity field in the bottom
panel of \reffig{fig:n2273_velfield}.

In general however, with non-negligible errors, any visualization
scheme that tries to reproduce the measurements exactly creates
artifacts whose reality is difficult to judge. The problem lies in the
fact that in 2D kinematics maps there is no standard way to visualize
error, which one is used to see in 1D kinematics profiles. For this
reason it becomes difficult to distinguish noise fluctuations from
real features in the data. A solution would be to use standard non-
parametric methods, like `smoothing splines' \citep{wha90} or `kernel
estimators' \citep{sco92}, to recover the kinematical field and
display the maps, taking measurements errors into account. This
assumes of course that the intrinsic 2D field under study is smooth on
the scale of a bin. If this is not the case, there is no way one can
tell it from the data: binning is in fact only applied when noise
dominates over the small spatial scales, and the fine details are
already lost.

The recovery of the intrinsic velocity field is however beyond the
scope of the present paper. In 2D like in 1D, binning is generally
simply used to obtain a reliable measure of some quantity inside a
given spatial region. The measure is then compared with models (e.g.\
dynamical ones), which are evaluated exactly inside the same spatial
region as the real observation. As long as the assumption in the model
are correct, and model and data are treated in the same way, one does
not have to worry about what happens inside each bin.

Similar tests of binning IFS data were performed on a number of
objects for which data obtained with \sauron\ were available. In
addition we also performed 2D binning of galaxy images. The results
presented here for NGC~2273 are representative of all the cases we
tried.

\subsection{Availability}

Software implementing in IDL\footnote{http://www.rsinc.com/} the
method described in this section is available from\\
http://www.strw.leidenuniv.nl/$\sim$mcappell/idl/.\\ In addition to
the desired target \SN, only four columns of numbers are required as
input: the coordinates $(x_i,y_i)$ of each pixel and the corresponding
signal $\mathcal{S}_i$ and noise $\mathcal{N}_i$. In output to each
pixel a bin number will be assigned.

The above IDL implementation took 3.5 s on a 1 GHz PC to generate the
2D-binning shown in \reffig{fig:bubbles-final}, which is composed of
3107 pixels grouped into 335 bins. The computation time scales roughly
with the number of pixels to bin. The memory needed by the algorithm
is comparable to that required to store the pixels themselves.

\section{Conclusions}
\label{sec:concl}

The problem of adaptively binning 2D data (e.g.\ spatial elements of
IFS or imaging data) to a constant \SN{} per bin has been
analyzed in detail. The goals towards which an optimal algorithm
should tend have been presented, and the limitations of different
approaches, making innovative use of the Voronoi tessellations, have
been discussed. Finally a robust algorithm that solves the 2D-binning
problem in an optimal way has been described. The different methods have
been applied to the binning and to the extraction of the stellar
kinematics from the \sauron{} data of the barred Sa galaxy
NGC~2273.

Binning is invariably used to analyze 1D (e.g.\ long-slit)
spectroscopic observations. We believe adaptive 2D-binning should
become common practice for the analysis of 2D data too (in particular
spectral data). This method is being used systematically in the
analysis of the IFS data obtained by the \sauron{} project.

Our binning methods can be naturally extended to three dimensions. The
CVT binning described here can also be used with discrete datasets,
like e.g.\ stellar proper motions observations for globular clusters,
or datasets coming from N-body simulations \citep[e.g.][]{sch00}, or
X-ray data.

\section*{Acknowledgments}

MC's research was supported through a European Space Agency external
fellowship. YC's research was supported through a European Community
Marie Curie Fellowship. We thank Tim de Zeeuw and Eric Emsellem for
helpful comments on the manuscript and Gijs Verdoes Kleijn for lively
discussions on 2D-binning. We are indebted to Qiang Du for
clarifications regarding the Gersho conjecture. We thank the \sauron{}
team for providing the data of NGC~2273 and for fruitful discussions
on the method.

\appendix

\section{Connection with the mesh generation problem}
\label{sec:delaunay}

Closely related to the VT, is the Delaunay triangulation (in 2D), that can
be obtained from the tessellation by connecting each generator to the
generators of the adjacent Voronoi regions. The Delaunay triangulations are
widely used in various applications such as Finite Elements methods and
interpolation schemes, due to the fact that, once the set of generators has
been defined, they optimize the geometric quality of the resulting
triangulation.

\begin{figure}
  \plotone{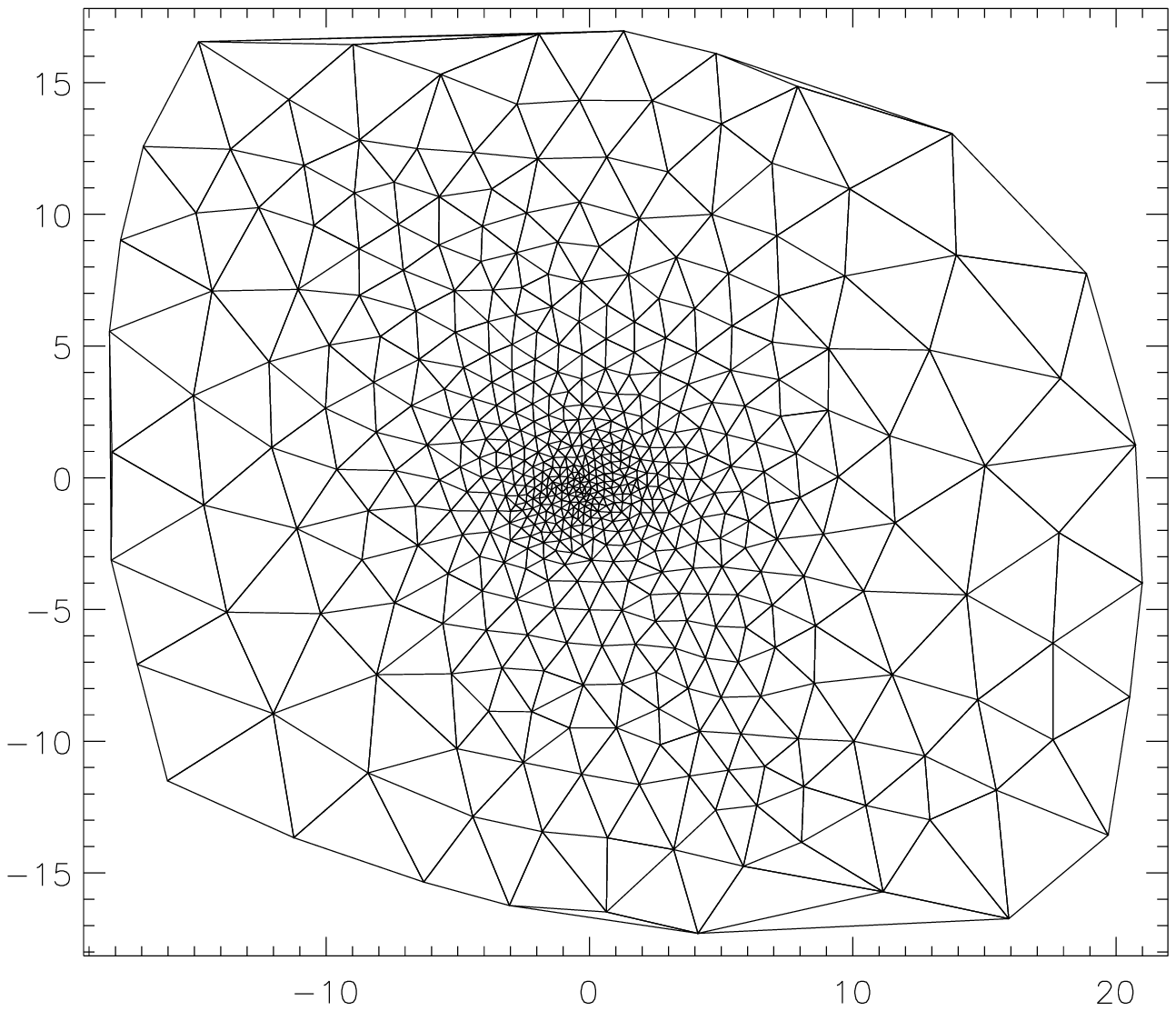}
  \caption{Delaunay triangulation computed from the generators of the
    VT shown in \reffig{fig:n2273-cvt}.}
  \label{fig:delaunay}
\end{figure}

A complementary way of looking at the regularity of the grid obtained
with the CVT consists of constructing the Delaunay triangulation of
the generators found by Lloyd's algorithm. This triangulation can be
easily computed by connecting with each other all generators of
adjacent bins and is represented in \reffig{fig:delaunay} for the CVT
of \reffig{fig:n2273-cvt}. The triangulation obtained is composed of
triangles whose shape is asymptotically close to equilateral, and
indeed it is easy to understand that hexagonal Voronoi regions
correspond to equilateral triangles of the Delaunay triangulation
computed from the same set of generators.

This fact makes obvious the close connection, in the continuum case,
between the problem of generating an optimal adaptive 2D-binning of a given
region and that of constructing an optimal triangulation of the same
region. These are in fact just two different ways of looking at the same
problem. Ignoring boundary effects, CVT are an extremely simple but
powerful tool for the construction of an unstructured mesh, and the
obtained triangulations are of quality comparable to the most advanced and
complex mesh generation methods. The $\mathcal{A}-\rho$ relation between
the size of CVT bins and the density, which derives from Gersho
conjecture, makes it possible to adapt the size of the mesh to any desired
distribution, while still preserving the equilateral characteristic of the
generated mesh.

\label{lastpage}
\end{document}